\documentclass[aps,prb]{revtex4}
\usepackage{bm}
\usepackage{amsmath}
\usepackage{color}
\usepackage{graphicx}
\usepackage{amssymb}
\usepackage[shortlabels]{enumitem}
\DeclareMathAlphabet{\mathbbmsl}{U}{bbm}{m}{sl}
\makeatletter
\newsavebox{\@brx}
\newcommand{\llangle}[1][]{\savebox{\@brx}{\(\m@th{#1\langle}\)}%
	\mathopen{\copy\@brx\kern-0.5\wd\@brx\usebox{\@brx}}}
\newcommand{\rrangle}[1][]{\savebox{\@brx}{\(\m@th{#1\rangle}\)}%
	\mathclose{\copy\@brx\kern-0.5\wd\@brx\usebox{\@brx}}}
\makeatother

\begin{document}
\draft
  \title{Floquet Control of Electron and Exciton Transport in Kekul\'e-Distorted Graphene}

\author{Sita Kandel$^{1,2}$ and Godfrey Gumbs$^{1,2}$ }
 \address{$^1$Department of Physics, Hunter College, City University of New York, 695 Park Avenue, New York, NY 10065 USA}
\address{$^{2}$The Graduate School and University Center, The City University of New York,    New York, NY 10016, USA}

\date{\today}

\begin{abstract}

This work investigates the Floquet dynamics of electrons and excitons (particle-hole pairs) in a Dirac material referred to as Kekul\'e-distorted graphene.  Specifically, we examine the role played by a high frequency driving electromagnetic field on the tunneling and blocking by a potential barrier  on both the charged single particles as well as the neutral composite particles. We demonstrate that  the small effective masses of the electron and hole for the energy spectrum of this Kekul\'e distorted graphene leads to practically almost perfect  transmission across a  symmetric potential barrier for any angle of incidence  of impinging excitons. However, this unexpected Klein paradox for excitons does not hold for the single-particle electrons. The reduced  total transmission of electron due to Kekul\'e distortion  is more  suppressed due to irradiation. Additionally, we calculate and investigate the exciton binding energy since the quantum tunneling of a bound electron-hole pair across a potential barrier is governed by its  mass  measured in the center of mass and binding energy of the composite pair.  Thus, irradiation with circularly polarized light   fundamentally modifies exciton formation, coherence and transport properties, thereby producing  unusual topological behaviors.  These behaviors are  unlike conventional Dirac materials.  Possible technical applications of the results arising from our investigation include valleytronics due to the folding of the valleys, thereby making intervalley coupling feasible.  Other practical applications include optoelectronics due to Floquet tuning of energy spectrum and transport properties. 
\end{abstract}

\maketitle

\medskip

\medskip


\section{Introduction}
 \label{sec1}
In recent times, spatially inhomogeneous two dimensional (2D) materials have been attracting significant scrutiny due to their potential usage to display novel physical phenomena \cite{deep, mac, naumis, chunmei, thinga,  peng}. Kekulé-distorted graphene \cite{bastos, gamayun, andrade, saul, mojaro, herera, bao} is one of the most   galvanizing materials among these low-dimensional materials. With arbitrary periodic perturbation on Kekulé-distorted graphen, the two nonequivalent Dirac cones at $K^{\pm}$ which are subject to time-reversal symmetry  at the corners of a corresponding hexagonal Brillouin zone like graphene  are folded at the center $\Gamma$ point of the new hexagonal superlattice Brillouin zone,  resulting in either a gap (Kek-O) or the superposition of two cones with different Fermi velocities (Kek-Y)\cite{bastos, gamayun, andrade}. The Kekul\'e-phase was first reported in graphene sheets epitaxially grown over copper substrates\cite{brown} and also in monolayer graphene on top of silicon oxide\cite{koo}. After the experimental realization, several works have been reported exploring the physical\cite{bastos,koo} electronic\cite{gamayun}, optical\cite{herera, mojaro}, transport\cite{juan} and several more intriguing properties\cite{arraga, mireles, chang, bahrami} including spin-orbit interaction, \cite{koltai} as well as phonon\cite{Duan} or magnon\cite{panta, mouls} excitations. Recently, direct experimental evidence for chiral symmetry breaking in  a Li-intercalated Kekul\'e-ordered  graphene from both electron spectroscopic and microscopic measurements have been reported. \cite{bao}
\medskip
\par
In this work, we consider the Y-shaped Kek lattice as schematically exhibited in Fig. \ref{FIG:1}, where one of  six carbon atoms in each superlattice unit cell display three shorter nearest-neighbor bond (green lines) shaped like the letter Y. The  superlattice unit cell  of the Kekul\'e distorted graphene is $\sqrt{3}\times\sqrt{3}$ larger than the graphene unit cell. The structure preserves inversion symmetry and two chiral valleys coupled together to form massless Dirac bands leading to vanishing valley degeneracy. We study the valley coupled tunneling of electrons in such Y-shaped Kek structure calculating  the probability currents explicitly for both reflection and transmission of electrons across a potential barrier of finite height and width. By matching boundary conditions at the interface of this barrier, we acquire valley resolved reflection and transmission coefficients as functions of the incident kinetic energy and the angle of incidence, as well as the barrier height and width for chosen Kekul\'e parameter. Our present study reveals that the intervalley coupling enhances the perfect transmission around normal incidence for small barrier width. However,  the resonant tunneling  is supressesd when the barrier width is increased.    We have been  motivated to investigate the tunneling behavior by opening a gap between the conduction and valence bands as well as between two folded cone in such a structure. 

\medskip
\par
There are many ways of modifying the band structure of Dirac like electronic system. Among them, manipulating the electronic properties by using  high-frequency dressing field  which is called Floquet engineering has become a very powerful and efficient technique nowadays in the era of 2D materials\cite{kibis, iurov2, iurov4, jong, albetro, johnson} although it is not a new concept in quantum physics. The field of Floquet engineering -where matter is coherently driven by periodic temporal perturbations- was initiated by Fistul and Efetov who proposed that a dynamically opened gap, generated by irradiating graphene with an electromagnetic field, could be employed to control Klein tunneling in a p-n junction.\cite{fistul, efetov}  Later, Oka and Aoki recognized the nontrivial nature of this gap and predicted a light induced topological anomalous Hall state associated with an induced magnetic field  due to symmetry breaking.\cite{oka} Merboldt et al.\cite{merbolt} recently revealed an experimental framework for implementing Floquet-engineering protocols in metallic and semimetallic platforms with potentially nontrivial topological characteristics. These developments motivate the exploration of periodically driven 2D semimetals such as graphene. Here, we present the effect due to high frequency circularly polarized light irradiation on band structure and tunneling behavior. The result shows that a small difference is  established between the valleys and  a small gap is opened between the valence and conduction energy bands. This gap opening irradiation suppresses the tunneling of electrons but still preserves the perfect transmission for head-on collision.

\medskip
\par
In gapped Kek-Y graphene, the spatially separated electrons and holes form a bound quasi particle usually referred to as excitons. Study of quantum mechanical tunneling of composite particles dates back to around the 1960's by Zakhariev et al.\cite{zakharav,amirkhanov}. Ever since that time, it has become an active area of research investigation. For example, in 1964 Saito and Kayanuma\cite{saito} investigated the  resonant tunneling of a pair of bound particles through a single potential barrier, Kavka et al. \cite{kav1, kav2, kav3}  reported the time dependent formalism of molecular tunneling, of excitons through quantum heterostructure barrier and tunneling of molecules in three dimensions. Exciton-assisted electron tunneling in vander Waals heterostructures consisting of graphene and gold electrodes separated by hexagonal boron nitride with an adjacent TMD  (transition-metal dichalcogenide) monolayer was  recently published in ``nature materials”. \cite{lujun} Similarly, exciton-assisted resonant tunneling has been observed in conventional semiconductor quantum wells.\cite{cao1, cao2} In this respect, tunneling of excitons in gapped Kek-Y graphene across a finite width square potential  barrier of equal strength  as well as delta potential of inequivalent strength for electrons and holes becomes an interesting problem to include in this paper. We have merged together in this work  the exciton binding energy and   the  reflection and transmission probability of  an exciton since the survival of an exciton after it impinges the potential barrier is relevant in this calculation. 
\medskip
\par
The   remainder of this paper is organized in the  following way.  In  Sec.\  \ref{sec2},   we introduce a low-energy Hamiltonian for Kek-Y graphene  and briefly  review some essential properties resulting  from the energy dispersion and the corresponding electronic eigenfunctions in the absence of irradiation.  Section \ref{sec3} is devoted to deriving  an effective Hamiltonian  which has the same stroboscopic effects as the externally applied time-dependent periodic potential  employed   to irradiate  the system  with elliptically polarized light.  We also  analyze the electron-photon dressed states under this irradiation.   In Sec. \  \ref{sec4}(A),  we calculate the probability current, transmission and reflection probability of a particle incident on the potential barrier and observe the valley resolved Klein tunneling in the absence of irradiation.  With help from the derived transmission probability, we  obtain  numerically the quantum conductivity through a monolayer of Kek -Y distorted graphene. Our results for  the effect of circularly polarized irradiation on transmission and reflection  are presented  in Sec. \ref{sec4}(B). Tunneling of excitons on irradiated gap-induced Kek-Y graphene is calculated and analyzed in Sec. \ \ref{sec5}.  The final conclusions of our presented results and  outlook are presented in Sec.\  \ref{sec6} and  the Appendix includes the steps and lengthy expressions behind some complicated mathematical calculations.     

\medskip
\par

\section{Low-energy  Hamiltonian  and ebergy eigenstates for Kek-Y graphene}
\label{sec2}

\begin{figure} 
\centering
\includegraphics[width=0.3\textwidth]{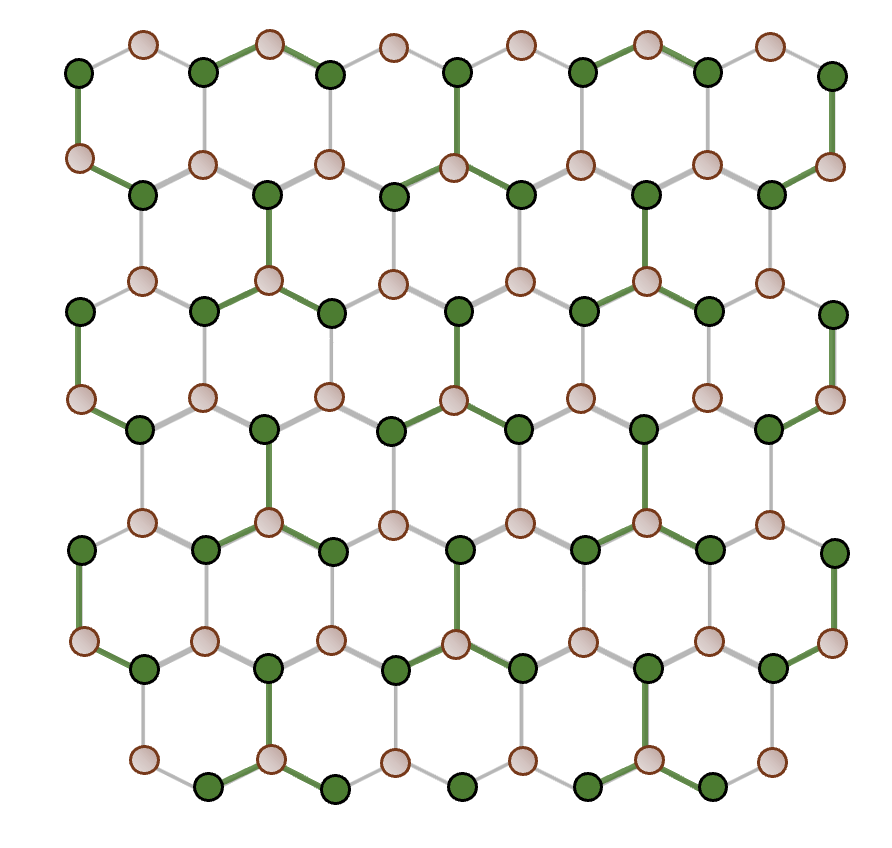}
\includegraphics[width=0.3\textwidth]{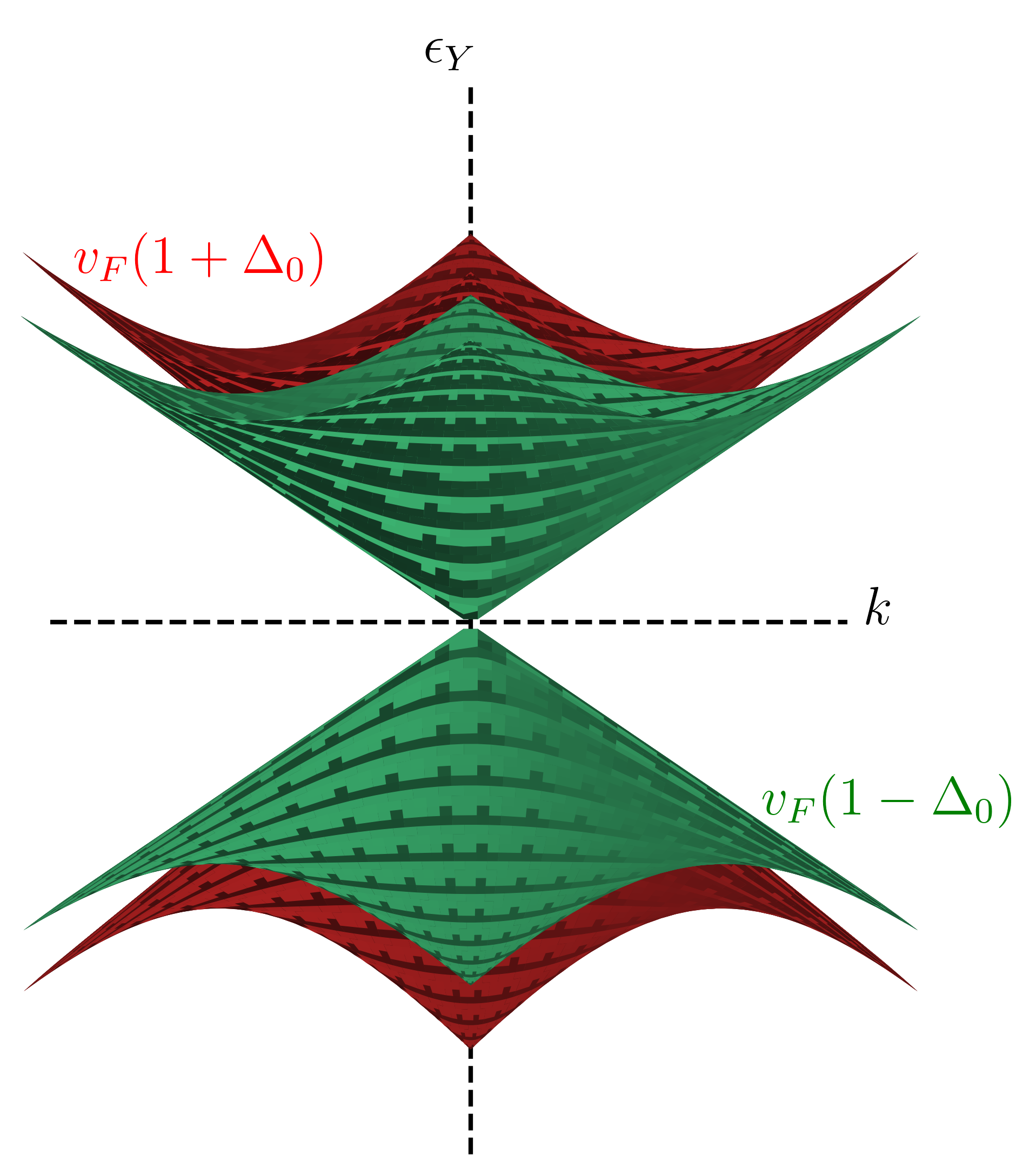}
\caption{(Color online) Kekul\'e distorted honeycomb lattice (left) and pictorial representation of two Dirac cones (right) of  Kek-Y distorted graphene .}
\label{FIG:1}
\end{figure}

First, to establish notation, we briefly review the electronic band structure and eigenfunctions in Kek-Y distorted graphene. For this type of distortion, the Dirac cones will fold on each other, as shown in Fig.\,\ref{FIG:1}, which results in two Dirac cones with one buried inside the other with no bandgap, but with different Fermi velocities. 
\medskip

The low-energy Hamiltonian for Kek-Y graphene can be expressed as\cite{GG, mojaro}

\begin{equation}
\label{mainH}
\hat{\cal{H}}_Y (\mbox{\boldmath${k}$}\,\vert\,\Delta_0) = \hbar \upsilon_F   \hat{\cal{T}}_0^{(2)} \otimes \left(\mbox{\boldmath${k}$}\cdot \hat{\mbox{\boldmath${\Sigma}$}}^{(2)} \right) 
+\hbar \upsilon_F \, \Delta_0\, \left(\mbox{\boldmath${k}$}\cdot\hat{\mbox{\boldmath${\cal{T}}$}}^{(2)} \right) \, \otimes \hat{\Sigma}_0^{(2)}\  ,
\end{equation}
where $\mbox{\boldmath${k}$}=\{k_x,\,k_y\}$ is a wave vector of electrons,  $\upsilon_F$ is the Fermi velocity in regular unstrained graphene and $\Delta_0$ is a dimensionless strain-induced coupling parameter. As $\Delta_0 \rightarrow 0$, we recover the  Dirac Hamiltonian  for pristine graphene for unfolded $\mbox{\boldmath${\tau_+}$}$ and $\mbox{\boldmath${\tau_-}$}$ valleys, {\em i.e.\/},  retaining only the first term on the right hand side of Eq.\,\eqref{mainH}. We indicate that Eq.\,\eqref{mainH} is only valid for  a small strain-coupling parameter $\Delta_0$, and it may be modified as the strength  of the strain is increased. However, in  most cases,  $\Delta_0$ is assumed small in order to retain the graphene signature of the material. Here, we choose $\Delta_0 \backsim 0.1$ and $0.2$  in our calculations but derive a generalized model by assuming a substantial difference between two Fermi velocities $\upsilon_F\,(1\pm\Delta_0)$.  Additionally,

\medskip 
\par

\begin{eqnarray}
\label{six}
\hat{\Sigma}_x^{(2)} &= & \hat{\cal{T}}_x^{(2)} = \left(
\begin{array}{cc}
 0 & 1 \\
 1 & 0 
\end{array}
\right)\ ,
\end{eqnarray}
\begin{eqnarray}
\label{siy}
\hat{\Sigma}_y^{(2)} &=&  \hat{\cal{T}}_y^{(2)} =  \left(
\begin{array}{cc}
 0 & - i \\
 i & 0 
\end{array}
\right)\ ,
\end{eqnarray}
which act in the spin- and pseudo-spin subspaces, respectively. Correspondingly, $\hat{\Sigma}_0^{(2)}$ and $\hat{\cal{T}}_0^{(2)}$ in Eq.\,\eqref{mainH} are two $2 \times 2$  unit matrices in the spin- and pseudo-spin subspaces, written as 

\begin{eqnarray}
\label{si0}
\hat{\Sigma}_0^{(2)} &=&  \hat{\cal{T}}_0^{(2)} =  \left(
\begin{array}{cc}
 1 & 0 \\
 0 & 1 
\end{array}
\right)\ .
\end{eqnarray}
\medskip 

Equation\,\eqref{mainH} can be expressed in a simple matrix form

\begin{equation}
\hat{\cal{H}}_Y (\mbox{\boldmath${k}$}\,\vert\,\Delta_0) = \hbar \upsilon_F \, \left( \begin{array}{c  c }
\mbox{\boldmath${k}$}\cdot \hat{\mbox{\boldmath${\Sigma}$}}^{(2)} &  \Delta_0 \, (k_x - i k_y) \, \hat{\Sigma}_0^{(2)} \\
\Delta_0 \, (k_x + i k_y) \, \hat{\Sigma}_0^{(2)} & \mbox{\boldmath${k}$}\cdot \hat{\mbox{\boldmath${\Sigma}$}}^{(2)} 
\end{array}
\right)\ ,
\label{oph}
\end{equation}
or explicitly as 

\begin{equation}
\label{finalHam}
\hat{\cal{H}}_Y (\mbox{\boldmath${k}$}\,\vert\,\Delta_0) = \hbar v_F \, \left( \begin{array}{c  c | c  c}
0 & k_x - i k_y & \Delta_0 (k_x - i k_y) & 0 \\
k_x + i k_y & 0 & 0 & \Delta_0 (k_x - i k_y)  \\
\hline
\Delta_0 (k_x + i k_y) & 0 & 0 & k_x - i k_y \\
0  & \Delta_0 (k_x + i k_y) & k_x + i k_y & 0 
\end{array}
\right)\ .
\end{equation}
\medskip 
\par

\begin{figure} 
\centering
\includegraphics[width=0.5\textwidth]{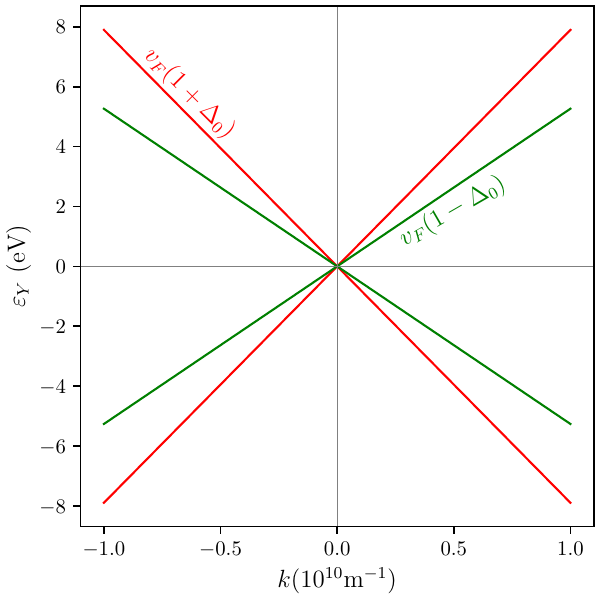}
\caption{(Color online)  Energy dispersions $\varepsilon_Y(s,\tau\,\vert\,k, \Delta_0)$ for Kek-Y graphene represented by two inequivalent Dirac cones with different Fermi velocities $\upsilon_{F,1}$ and $\upsilon_{F,2}$ corresponding to $\upsilon_F (1 \pm \Delta_0)$ .}
\label{FIG:2}
\end{figure}

The energy dispersions of Kekul\'e-Y graphene are simply obtained as eigenvalues of the matrix defined in Eq.\,\eqref{finalHam}, yielding

\begin{equation}
\varepsilon_Y(s, \tau\,\vert\,k, \Delta_0) = s \hbar \upsilon_F \, ( 1+ \tau \, \Delta_0) \, k \ ,
\label{disper} 
\end{equation}
which represent two Dirac cones with different Fermi velocities $\upsilon_F (1 \pm \Delta_0)$, as displayed in Fig.\,\ref{FIG:2}, where $s=\pm 1$ is a band index while $\tau=\pm 1$ corresponds to two different Fermi velocities due to strain coupling. These two types of subbands are often referred to as ``fast'' ($\tau=+1$) and ``slow'' ($\tau=-1$) Dirac cones, and importantly, photo-induced electronic transitions are allowed between these two. 
\medskip 
\par

In correspondence with energy dispersions in Eq.\,\eqref{disper}, the normalized wave functions are calculated as

\begin{equation}
\label{wave1}
\Psi_Y(s,\tau\,\vert\,\mbox{\boldmath${k}$},\Delta_0 ;{\bf r}) = \frac{1}{2} \, \left(
\begin{array}{c}
 \text{e}^{- i \theta_{\bf k}} \\
s\\ 
s \tau\\
\tau \, \text{e}^{ i \theta_{\bf k}} 
\end{array}
\right) 
\frac{e^{ik_xx+ik_yy}}{\sqrt{{\cal A}}}, 
\  ,
\end{equation}
where $\theta_{\bf k} = \tan^{-1}(k_y/k_x)$ is the angle associated with the electron wave vector $\mbox{\boldmath${k}$}$ with respect to the  $x$ axis. The wave function in Eq.\,\eqref{wave1} consists of two valley's $(\tau_+, \tau_-)$ sub-spinors corresponding to standard graphene eigenstates and are independent of strain applied to the system.  The two valleys are related by time-reversal symmetry $(\tau_+ = -\tau_-)$.  Also,  $A$ is a normalization area. For which, Berry’s phase, 
defined as  a phase difference acquired over the course of a cycle, when a system is subjected to cyclic adiabatic processes,  is 
\begin{equation}
\label{Berry1}
\phi_B=-i\oint  \frac{1}{4}
 \left(
 \text{e}^{ i \theta_{\bf k}} \  \  \
s \  \  \ 
s \tau  \  \  \
\tau \, \text{e}^{- i \theta_{\bf k}}  \right) 
\nabla_{{\bf k}}
\left(
\begin{array}{c}
\, \text{e}^{- i \theta_{\bf k}} \\
s\, \\ 
s  \tau \\
\tau  \text{e}^{ i \theta_{\bf k}} 
\end{array}
\right) d{\bf k}
\ .
\end{equation}
Carrying out the derivative gives

\begin{equation}
\label{Berry2}
\phi_B=  \frac{-i}{4}\oint
\left(
\text{e}^{ i \theta_{\bf k}} \  \  \
s \  \  \ 
s \tau  \  \  \
\tau \, \text{e}^{- i \theta_{\bf k}} \right)
\left(
\begin{array}{c}
-i  \,\text{e}^{- i \theta_{\bf k}} \nabla_{{\bf k}}\theta_{{\bf k}}\\
0 \\ 
0\\
i \tau \, \text{e}^{i \theta_{\bf k}}  \nabla_{{\bf k}}\theta_{{\bf k}} 

\end{array}
\right) d{\bf k} \  
\end{equation}
which  reduces to

\begin{equation}
\phi_B=  \frac{-i}{4}
(-i + i ) \oint  d\theta_{{\bf  k}}=0\    .
\label{Berry3}
\end{equation}
Therefore,  the Berry phase is $0 $ for the non-irradiated Kekul\'e distorted graphene as it is for the dice lattice but unlike pristine monolayer graphene for which the Berry phase has a magnitude of $\pi$.  For bilayer graphene, it is $2\pi$ and that for $\alpha - T_3$  varies between 0 to $\pi$.  A Berry phase of  $\pi $ simply applies to a pure relativistic Hamiltonian with both positive and negative energy states and that of zero represents non-relativistic Hamiltonian with only positive states and quadratic dispersion of energy with momentum. This is no longer is the case for  Kek-Y distorted graphene for the zero value of Berry phase  where low-energy dispersions are linear with both positive and negative states. It  is associated with the equal and  opposite chirality of two folded valleys at the same K-point of momentum space. This result in Eq.\ (\ref{Berry3}) is true for the conduction and valence bands as well as $\tau_+$ and $\tau_-$ valleys.

\medskip
\par

\section{Kek-Y Distorted Graphene Under Elliptically Polarized Electromagnetic Radiation }
 \label{sec3}

In the presence of an external perturbation due to an external electromagnetic wave, we can write the total Hamiltonian as  $\hat{H}^{T}(t)  = \hat{\cal{H}}_Y (\mbox{\boldmath${k}$}\,\vert\,\Delta_0)+\hat{V}_A^{(E)}(t) $ where

\begin{eqnarray}
\label{Wt}
\hat{V}_A^{(E)}(t)
= \left(
\begin{array}{cc}
 -e\upsilon_F  {\bf \sigma}  \cdot {\bf A} & \tilde{\Delta} W(t)\\
 \tilde{\Delta}^\ast W^\dag (t)& -e\upsilon_F  {\bf \sigma}  \cdot {\bf A}
\end{array}
\right)\ ,
\end{eqnarray}
 where $W(t)=-e\upsilon_F (A_x-iA_y)$ is expressed in terms of the vector potential ${\bf A}(t)$  and  $\hat{V}_A^{(E)}(t)$ is periodic with period $2\pi/\Omega$. For an elliptical polarization,  $\mbox{\boldmath$A$}^{(E)}(t)$ is given  by

\begin{equation}
\label{ellipA}
\mbox{\boldmath$A$}^{(E)}(t) =
\left[  \begin{array}{c}
          A^{(E)}_x (t) \\
          A^{(E)}_y (t)
        \end{array}
\right] = \frac{E_0}{\Omega} \left[
\begin{array}{c}
\cos \Theta_p \cos (\Omega t) - \beta \, \sin \Theta_p \sin (\Omega t) \\
\sin \Theta_p \cos (\Omega t) + \beta \,\cos \Theta_p \sin (\Omega t)
\end{array}
\right] \, 
\end{equation}
where $E_0$ is the electric-field strength,  $\Theta_p$ represents the polarization angle of the optical field which is  measured with respect to the normal to the surface and $\theta_k$ is the  phase angle of electron wave vector which is measured with respect to the $x-$axis. This Hamiltonian is periodic in time and follows the time-dependent Schr\"odinger equation given by 

\begin{equation}
i\hbar\frac{\partial  \Psi (\bf {k}, \bf {t})}{\partial t} =  \hat{H}^{T}(t) \Psi (\bf {k}, \bf {t}).
\label{SCHE}
\end{equation}

\begin{figure} 
\centering
\includegraphics[width=0.45\textwidth]{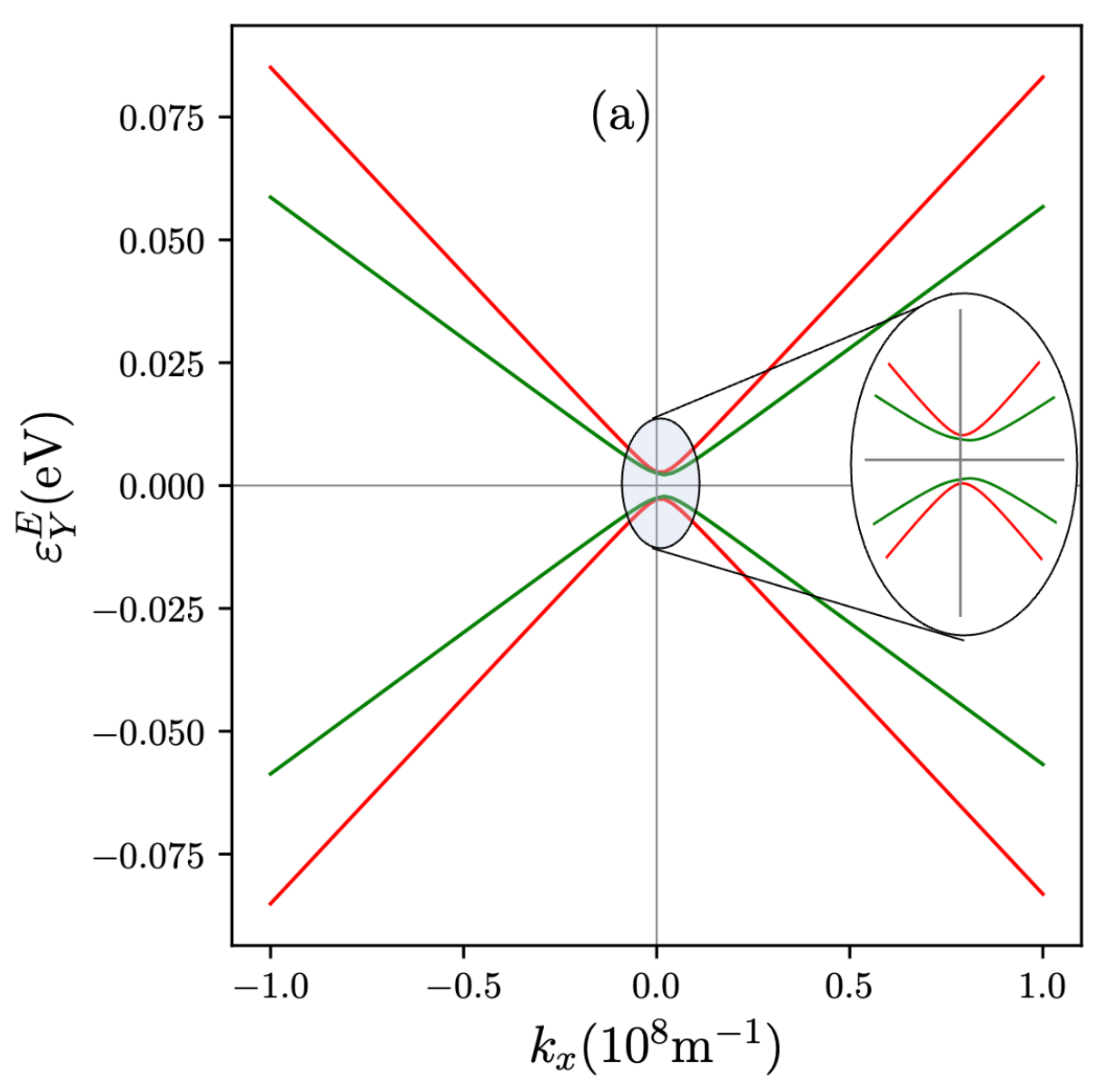}
\includegraphics[width=0.45\textwidth]{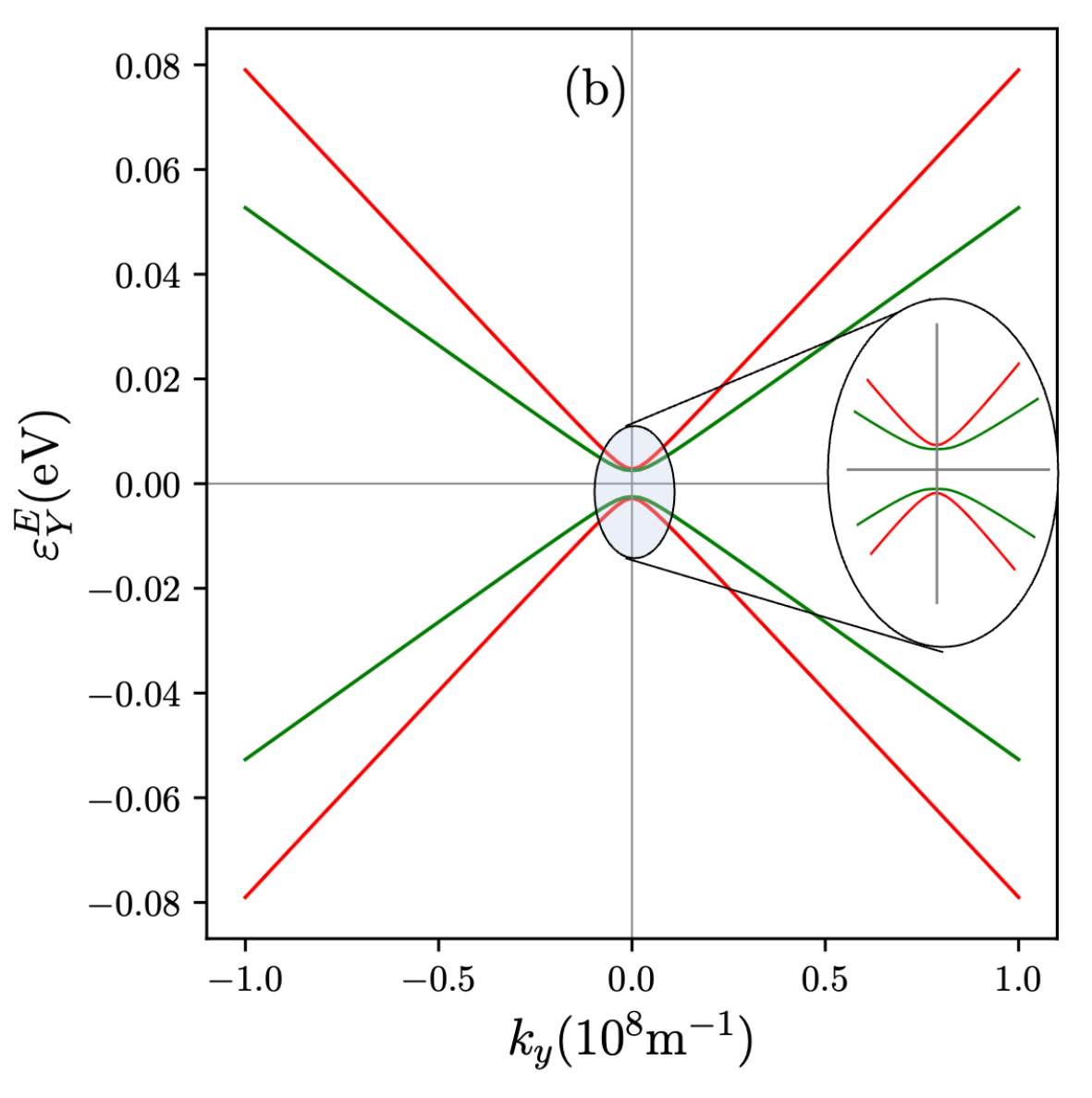}
\caption{(Color online) Energy dispersions $\varepsilon_Y(s,\tau\,\vert\,k, \Delta_0)$ for Kek-Y graphene under normal incident circularly polarized  irradiation  (i.e., $\theta_p =0$, $\beta = 1$)  represented by two inequivalent Dirac cones with different Fermi velocities $\upsilon_{F,1}$ and $\upsilon_{F,2}$. Plots in  (a) are along $ k_x$  and  plots in (b) are along $k_y$. The energy bands are anisotropic along two perpendicular momentum directions .  In each plot, the dispersion close to the Dirac points are focused,  which shows the opening of band gaps and vertical separation of the valleys.  These results are for light irradiation with frequency  of 100 THz, $\Delta_0 =0.2$ and $\tilde{\zeta} = 0.1$. }
\label{FIG:3} 
\end{figure}

\noindent
Based on the Floquet theorem, the solution to Eq.\, \eqref{SCHE} satisfies the periodicity condition with
\begin{equation} 
 \Psi (\bf{k}, \bf{t+T})=\text{e}^{-i H_{eff}T/ \hbar} \Psi (\bf{k}, \bf{t})  \  ,
 \label{PSI}
 \end{equation}
 where $H_{eff}$ is the Floquet effective Hamiltonian, which we obtain by employing the Floquet-Magnus expansion for a high-frequency periodic field.  Now, for elleptically polarized light for $\beta \neq 0$, the Floquet effective Hamiltonian for Kek-Y phase of graphene is as follows 

\begin{equation}
H^{E}_{eff}(k, \Delta_0,\zeta)
   =\left(
 \begin{array}{cccc}
  -C (1+ \Delta_{0}^{2}) & \hbar \upsilon_f (k_x -i k_y) &\Delta_0  \hbar \upsilon_f (k_x - i k_y)&  0 \\
  \hbar \upsilon_f (k_x +i k_y) & C (1- \Delta_{0}^{2})   \, \   &  0 &  \Delta_0  \hbar \upsilon_f (k_x -i k_y) \\
 \Delta_0  \hbar \upsilon_f (k_x +i k_y) & 0 & -C (1- \Delta_{0}^{2})   &  \hbar \upsilon_f (k_x -i k_y)\\  
    0 & \Delta_0  \hbar \upsilon_f (k_x +i k_y)& \hbar \upsilon_f (k_x +i k_y)& C (1+ \Delta_{0}^{2})    \end{array}
 \right) \  ,
 \label{HAC}
\end{equation}
 where we  have introduced $\tilde{\zeta}=\frac{-e\upsilon_F E_0}{\hbar \Omega^2}$  and  $C= \left[\tilde{\zeta}^2 \beta \hbar \Omega- 2 \tilde{\zeta} \beta \left(\hbar \upsilon_f k_x  \cos\theta_p + \hbar \upsilon_f k_y  \sin\theta_p \right) \right]$.  

\medskip
\par
Solving the eigenvalue equation of this effective Hamiltonian, we obtained the quasiparticle energy spectrum for Kek-Y–distorted graphene under elliptically polarized electromagnetic radiation is

\begin{equation}
\varepsilon^{E}_Y(s,\tau\vert k, \Delta_0,\zeta) =  s \left\{ \sqrt{C^2 + \hbar^2 \upsilon^2_f k^2}  + \tau \Delta_{0}
\sqrt{C^2 \Delta^2_{0}+ \hbar^2 \upsilon^2_f k^2} 
\right\}.
\label{MEX}
\end{equation}

\medskip
\par

 This type of polarized external radiation determines whether a significant gap between the cones in the valance and conduction bands is opened up at the Dirac point. The band gap between the inner cones and the gap between the outer cones are respectively given by the following equation for  $\tau = \pm 1$:
 
 \begin{equation}
\varepsilon^{E}_{gY}(\tau,  \Delta_0,\zeta) = 2 \hbar \Omega \tilde{\zeta}^2 \beta(1 + \tau  \Delta_{0}^2) .
\end{equation}
 Additionally, the two cones are also separated by small energy $ 2 \hbar \Omega \tilde{\zeta}^2 \beta\Delta_{0}^2 $,  at $k \to 0$ in each band.
The corresponding normalized wave functions are calculated as

\begin{equation}
\label{wave1C}
\Psi^{E}_{Y}(s,\tau\,\vert\,\mbox{\boldmath${k}$},\Delta_0 ;{\bf r}) = \  \frac{1}{\sqrt{(\text{b}^{2}_{2}+1)(\text{b}^{2}_{3} +1)}} \, \left(
\begin{array}{c}
\text{b}_1  (s, \tau;{\bf k} ) \text{e}^{- i \theta_k} \\
\text{b}_2  (s, \tau;{\bf k} )  \\
\text{b}_3  (s,\tau;{\bf k} )  \\
\text{b}_4  (s,\tau;{\bf k} ) \text{e}^{ i \theta_k}        
\end{array}
\right)
\frac{e^{ik_x x +i k_y y}}{\sqrt{A}},
\    ,
\end{equation}
where

\begin{eqnarray}
\label{wave1CB}
\text{b}_1  (s,\tau;{\bf k} ) &=&  1\nonumber\\
\text{b}_2  (s,\tau;{\bf k} ) &=&  
 \frac{ C + s \sqrt{C^2 + \hbar^2 \upsilon^2_f (k^2_x+k^2_y) }}{ \sqrt{\hbar^2 \upsilon^2_f (k^2_x+k^2_y)}  }
\nonumber\\
\text{b}_3  (s, \tau;{\bf k} ) &=& \frac{ \sqrt{\hbar^2 \upsilon^2_f (k^2_x+k^2_y)}  }{ s \tau \sqrt{C^2\Delta_{0}^2 + \hbar^2 \upsilon^2_f (k^2_x+k^2_y)} - C\Delta_0}\nonumber\\
\text{b}_4  (s, \tau;{\bf k} ) &=&  \frac{ C + s_1 \sqrt{C^2 + \hbar^2 \upsilon^2_f (k^2_x+k^2_y) }}{ s  \tau \sqrt{C^2\Delta_{0}^2 + \hbar^2 \upsilon^2_f (k^2_x+k^2_y)} - C\Delta_0} \end{eqnarray}
\medskip
\par
The band structure appearing in Fig.\,\ref{FIG:3} is for circularly polarized light ($\beta=1$)  at normal incidence. The plotted spectrum clearly shows a small gap between the valence and conduction bands  along with a vertical gap between the two cone. Notably,  the bands are anisotropic along two perpendicular momentum directions. In order to include the anisotropy of the bands, the circularly polarized  light at normal incidence  is used in further calculation.

\section{Kek-Y valley-coupled quantum tunneling}
\label{sec4}
\begin{figure} 
\centering
\includegraphics[width=0.8\textwidth]{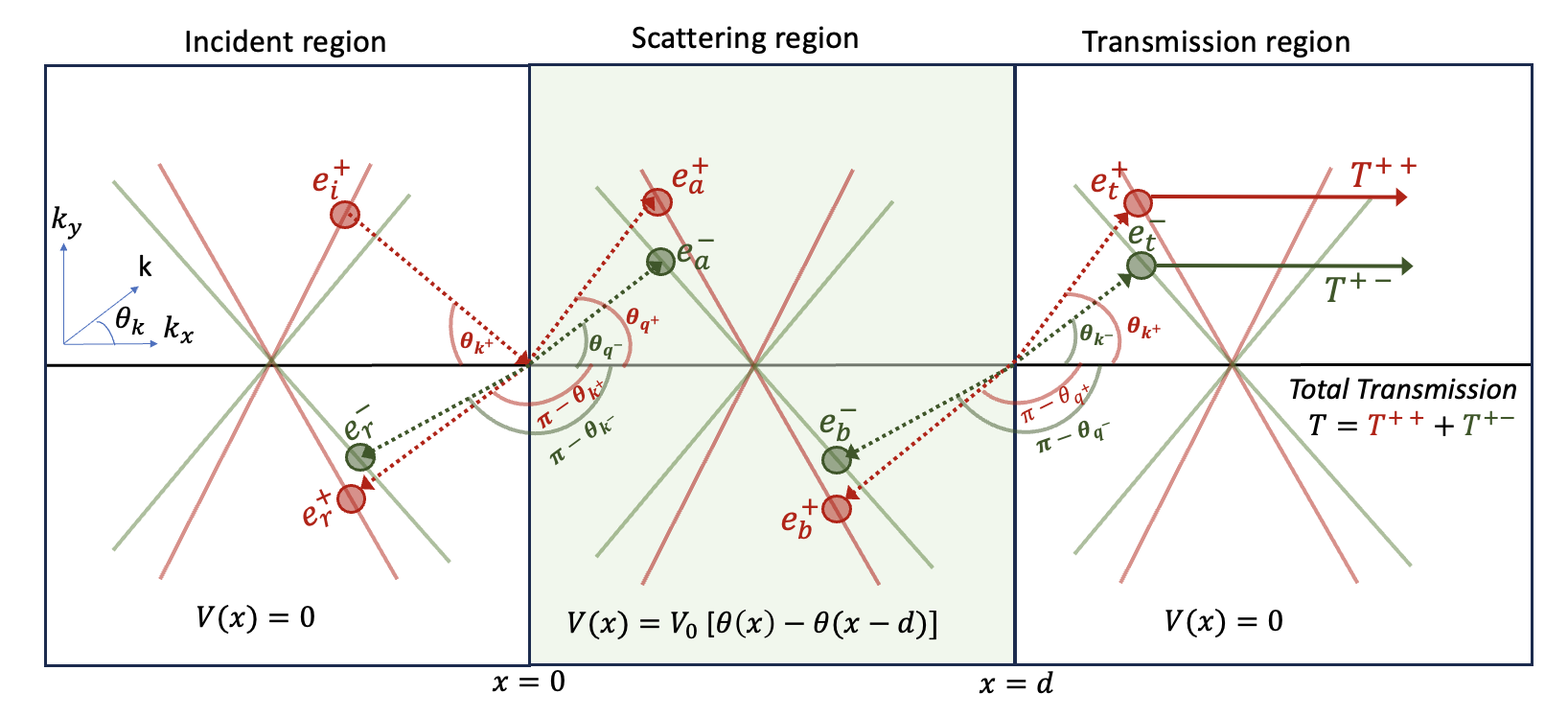}
\caption{(Color online) 
Schematics of electron transmission through potential barrier in Kek-Y graphene. The inner red and outer green cone represent the fast and slow cone or $\tau_+$ and $\tau_-$ valley respectively. The Dirac electron $e^+_i$ from $\tau_+$ valley in incident region is incident to the barrier at $x=0$ from left with incident angle $\theta_{k^+}$. This electron crosses the barrier of width $d$ and height $V_0$ and transmit to the transmission region as a $\tau_+$ valley electron $e^+_t$ with angle of transmission $\theta_{k^+}$ and probability $\cal T_{++}$ or as a $\tau_-$ valley electron $e^-_t$ with angle of transmission $\theta_{k^-}$ and probability $\cal T_{+-}$ such that the total probability of transmission $\cal T = \cal T_{++} + \cal T_{+-}$. $e^+_r, e^-_r$  are reflected electrons through the barrier with angle of reflection $\pi-\theta_{k^+}$ and $\pi-\theta_{k^-}$ as a $\tau_+$ valley and $\tau_-$ valley electrons respectively.  $e^+_a$ and $e^-_a$  are the forward moving electron at $x=0$ in scattering region  through $\tau_+$ valley and $\tau_-$ valley making angles $\theta_{q^+}$ and $\theta_{q^-}$ respectively. Similarly,  $e^+_b$ and $e^-_b$  are the backward moving electron at $x=d$ in scattering region  through $\tau_+$ valley and $\tau_-$ valley making angles $\pi-\theta_{q^+}$ and $\pi- \theta_{q^-}$ respectively.  }
\label{FIG:4}
\end{figure}

For pristine graphene, the two non-equivalent Dirac K points are separated by a large momentum distance suggesting that intervalley scattering is unlikely at low energies. However, in Kek-Y graphene, the two cones are folded at same symmetry point in the Brillouin zone, making it possible for intervalley transport between the cones. The intervalley transport ensures the valley degree of freedom in graphene in the presence of strain. Since the valley precision is spatially anisotropic depending on the spacial symmetry of the material, we assume the translational symmetry to keep the transverse momentum conserve and  focus our attention on the calculation of probability current for determining the transmission or reflection coefficient along the longitudinal direction.   For this, we  introduce a  barrier $V(x)= V_0 [\Theta(x)- \Theta(x-d)]$ of width $d$ and  height  $V_0 $ along the $x$ direction of the system. The schematic illustration of electron transmission for  Kek-Y distorted graphene is given in Fig.\, \ref{FIG:4}. In the presented figure, the outer green cone and the inner red cone represents the slow and fast cone as we explained in Sec.\ I. The symbols $e_i, e_r, $ and $e_t$ denote the incident, reflected and transmitted electron,  respectively. For reference, the beam of electrons in the conduction band of the inner fast cone is incident on the potential with incident angle $\theta_{k^+}$. From the edge of the potential  at $x=0$,  and $x=d$, the particle either reflects or transmits to the adjacent region. The particle from the inner cone of  the incident region can appear as the particle of the outer cone in  the transmission region.  It means that there is finite probability for intervalley scattering.  Therefore, the total probability of  transmission $T$ is the sum of the probability of transmission from the same incident cone $\tau^{++}$ and that of the other cone $\tau^{+-}$, which we  respectively called intravalley and intervalley transmission. 
\medskip
\par
 In general, the time-dependent Schr{\"o}dinger equation for the Hamiltoniam  $\hat{{\cal H}}$ in Eq.\, \eqref{finalHam} as well as the spinor-type wave function $|\mbox{\boldmath$\Psi$}>=[\psi_A\ \psi_B\ \psi_C\ \psi_D]^T$, can be written as $i\hbar\,\partial/\partial t\,|\mbox{\boldmath$\Psi$}>=\hat{{\cal H}}\,|\mbox{\boldmath$\Psi$}>$ from which we find that the probability density $\rho=<\mbox{\boldmath$\Psi$}|\mbox{\boldmath$\Psi$}>$ satisfies

\begin{equation}
i\hbar\frac{\partial\rho}{\partial t} =i\hbar\frac{\partial}{\partial t} <\mbox{\boldmath$\Psi$}|\mbox{\boldmath$\Psi$}>
= <\mbox{\boldmath$\Psi$}|\hat{{\cal H}}|\mbox{\boldmath$\Psi$}>
-\left(   
 \hat{{\cal H}}|\mbox{\boldmath$\Psi$}>\right)^\dag |\mbox{\boldmath$\Psi$}>  \   .
\label{GG1}
\end{equation}

\subsection{Tunneling and electrical conductivity without Irradiation}
After substituting the Hamiltonian given in Eq.\,\eqref{finalHam} into Eq.\,\eqref{GG1}, we see after a lengthy calculation, that the probability current density $\mbox{\boldmath$j$}=\{j_x,j_y\}$ satisfies the continuity equation  $\nabla\cdot\mbox{\boldmath$j$}+\partial\rho/\partial t =0$, and  translational invariance along the $y$ direction  further yields that   $j_x$ is  conserved which is expressed as

\begin{eqnarray}
j_x &=&v_F \, \left\{(\psi_A^\ast\psi_B+ \psi_B^\ast\psi_A+\psi_C^\ast\psi_D +\psi_D^\ast\psi_C)
+\Delta_0(\psi_A^\ast\psi_C+\psi_C^\ast\psi_A +\psi_B^\ast\psi_D+\psi_D^\ast\psi_B)
\right\}\  .
\label{JJ}
\end{eqnarray}
The terms $\propto \Delta_0$ clearly show the contributions to the current arising from the inter-cone coupling. The other terms are simply due to the graphene part of the Hamiltonian presented in Eq.\, \eqref{finalHam}.
\medskip
\par
We now consider the  three regions depicted schematically in Fig.\ \ref{FIG:4}.  Region (I) is the incident region where the particle has  kinetic energy $\varepsilon_k$ and zero potential energy.  The particle on the $\tau_+$ cone, having momentum $k^{+}$ incident from the left of the barrier of width $d$ and constant potential height $V_0$ with incident angle $\theta_k^{+}$ is  represented by  the wave function,

\begin{equation}
\label{wavet}
\Psi_Y^{(i)}(s,\vert\,\mbox{\boldmath${k}$};{\bf r}) = \frac{1}{2} \, \left(
\begin{array}{c}
\text{e}^{- i \theta_{\bf k^{+}}} \\
s\\ 
s\\
\, \text{e}^{ i \theta_{\bf k^{+}}} 
\end{array}
\right) 
\frac{e^{i|k^{+}_x|x+ik_yy}}{\sqrt{{\cal A}}}
\   .
\end{equation}
From Eq.\ \eqref{JJ}, the incident probability current density is calculated, 

\begin{eqnarray}
J_x^{(i)}(s,{\bf k}) &=& j_x^{(i)}(s,{\bf k}){\cal A}
= v_Fs \left[1+ \Delta_0 \right] \cos\theta_{{\bf k^{+}}} \ 
\label{J_i}
\end{eqnarray}
with $|\theta_{\bf k^{+}}|\leq\pi/2$  and $k^{+}_x = \frac{\varepsilon_k}{s\hbar\upsilon_F(1+\Delta_0)}\cos(\theta_k^{+})$, $k_y = \frac{\varepsilon_k}{s\hbar\upsilon_F(1+\Delta_0)}\sin(\theta_k^{+})$ and $s = $ sign($\varepsilon_k$).

Region (II) is the place where the particle has interaction with the potential $V_0$ and is launched  with initial momentum $q^{\tau}$ corresponding to  kinetic energy  $|\varepsilon_k -V_0|$. The electron from the $\tau_+$ valley in the incident region  travelled through the scattering region and gets transmitted into region (III)  which is referred to as the transmission region where the particle again has kinetic energy $\varepsilon_k$ as both the transmitted and reflected electron conserve energy.  When  transferred to the transmission region, it   either comes out to be a $\tau_+$ valley electron or $\tau_-$ valley electron with wave functions given by 

\begin{equation}
\label{wavet_p}
\Psi_Y^{(t_{+})}(s \,\vert\,\mbox{\boldmath${k}$};{\bf r}) = \frac{t_{+}}{2} \, \left(
\begin{array}{c}
\text{e}^{- i \theta_{\bf k^{+}}} \\
s\\ 
s\\
  \text{e}^{ i \theta_{\bf k^{+}}} 
\end{array}
\right) 
\frac{e^{i|k^{+}_x|x+ik_yy}}{\sqrt{{\cal A}}}\  , 
\end{equation} 
and 

\begin{equation}
\label{wavet_m}
\Psi_Y^{(t_{-})}(s \,\vert\,\mbox{\boldmath${k}$};{\bf r}) = \frac{t_{-}}{2} \, \left(
\begin{array}{c}
\text{e}^{- i \theta_{\bf k^{-}}} \\
s\\ 
-s\\
  -\text{e}^{ i \theta_{\bf k^{-}}} 
\end{array}
\right) 
\frac{e^{i|k^{-}_x|x+ik_yy}}{\sqrt{{\cal A}}}\  , 
\end{equation} 
where $\theta_{\bf k^{-}} = \tan^{-1}(k_y/k^{-}_x)$ is the angle associated with the electron wave vector ${\bf k^{-}} = \frac{\varepsilon_k}{s\hbar\upsilon_F(1-\Delta_0)}$, with respect to the  $x$ axis and $t_{\pm}$ are the transmission coefficients which are calculated by making use of matching boundary conditions. The transmitted probability current is the sum of the current through the $\tau_+$ valley and the $\tau_-$ valley. They are    
\begin{eqnarray}
J_x^{(t_+)}(s,{\bf  k^{+}}) &=& j_x^{(t_+)}(s,{\bf k^{+}}){\cal A}
= \upsilon_F s (1+\Delta_0)|t_+|^{2} \cos(\theta_k^{+})
\end{eqnarray} 
and 
\begin{eqnarray}
J_x^{(t_-)}(s,{\bf  k^{-}}) &=& j_x^{(t_-)}(s,{\bf k^{-}}){\cal A}
= \upsilon_F s (1-\Delta_0)|t_-|^{2} \cos(\theta_k^{-})\  .
 \end{eqnarray}    

Similarly, the electrons are reflected to the incident region with the wave functions,

\begin{equation}
\label{wavetr_p}
\Psi_Y^{(r_{+})}(s \,\vert\,\mbox{\boldmath${k}$};{\bf r}) = \frac{r_{+}}{2} \, \left(
\begin{array}{c}
-\text{e}^{ i \theta_{\bf k^{+}}} \\
s\\ 
s\\
 - \text{e}^{- i \theta_{\bf k^{+}}} 
\end{array}
\right) 
\frac{e^{-i|k^{+}_x|x+ik_yy}}{\sqrt{{\cal A}}}\ , 
\end{equation} 
and 

\begin{equation}
\label{wavet_m}
\Psi_Y^{(r_{-})}(s \,\vert\,\mbox{\boldmath${k}$};{\bf r}) = \frac{r_{-}}{2} \, \left(
\begin{array}{c}
-\text{e}^{ i \theta_{\bf k^{-}}} \\
s\\ 
-s\\
  \text{e}^{- i \theta_{\bf k^{-}}} 
\end{array}
\right) 
\frac{e^{-i|k^{-}_x|x+ik_yy}}{\sqrt{{\cal A}}}\  , 
\end{equation} 
where  $r_{\pm}$ are the reflection coefficients. The reflection probability currents are given by
\begin{eqnarray}
J_x^{(r_+)}(s,{\bf  k^{+}}) &=& j_x^{(r_+)}(s,{\bf k^{+}}){\cal A}
= -\upsilon_F s (1+\Delta_0)|r_+|^{2} \cos(\theta_k^{+})
\end{eqnarray} 
and 
\begin{eqnarray}
J_x^{(r_-)}(s,{\bf  k^{-}}) &=& j_x^{(r_-)}(s,{\bf k^{-}}){\cal A}
= -\upsilon_F s (1-\Delta_0)|r_-|^{2} \cos(\theta_k^{-})  \ .
\end{eqnarray} 

\medskip
\par
In the scattering region, the electron moving forward from the potential edge $x=0$ and moving backward from the edge $x=d$ from both $+$ and $-$ valleys has corresponding wave functions,

\begin{equation}
\label{wave_a_p}
\Psi_Y^{(\gamma_{+})}(s^{\prime} \,\vert\,\mbox{\boldmath${q}$};{\bf r}) = \frac{\gamma_{+}}{2} \, \left(
\begin{array}{c}
\text{e}^{ -i \theta_{\bf q^{+}}} \\
s^{\prime}\\ 
s^{\prime}\\
  \text{e}^{i \theta_{\bf q^{+}}} 
\end{array}
\right) 
\frac{e^{i|q^{+}_x|x+ik_yy}}{\sqrt{{\cal A}}}\  , 
\end{equation}

\begin{equation}
\label{wave_a_m}
\Psi_Y^{(\gamma_{-})}(s^{\prime} \,\vert\,\mbox{\boldmath${q}$};{\bf r}) = \frac{\gamma_{-}}{2} \, \left(
\begin{array}{c}
\text{e}^{ -i \theta_{\bf q^{-}}} \\
s^{\prime}\\ 
-s^{\prime}\\
  -\text{e}^{ i \theta_{\bf q^{-}}} 
\end{array}
\right) 
\frac{e^{i|q^{-}_x|x+ik_yy}}{\sqrt{{\cal A}}}\  , 
\end{equation} 

\begin{equation}
\label{wave_b_p}
\Psi_Y^{(\eta_{+})}(s^{\prime} \,\vert\,\mbox{\boldmath${q}$};{\bf r}) = \frac{\eta_{+}}{2} \, \left(
\begin{array}{c}
-\text{e}^{ i \theta_{\bf q^{+}}} \\
s^{\prime}\\ 
s^{\prime}\\
  -\text{e}^{-i \theta_{\bf q^{+}}} 
\end{array}
\right) 
\frac{e^{-i|q^{+}_x|x+ik_yy}}{\sqrt{{\cal A}}} 
\end{equation} 
 and 

\begin{equation}
\label{wave_b_m}
\Psi_Y^{(\eta_{-})}(s^{\prime} \,\vert\,\mbox{\boldmath${q}$};{\bf r}) = \frac{\eta_{-}}{2} \, \left(
\begin{array}{c}
-\text{e}^{ i \theta_{\bf q^{-}}} \\
s^{\prime}\\ 
-s^{\prime}\\
  \text{e}^{- i \theta_{\bf q^{-}}} 
\end{array}
\right) 
\frac{e^{-i|q^{-}_x|x+ik_yy}}{\sqrt{{\cal A}}}\  , 
\end{equation} 
where $\gamma_{\pm}$  and $\eta_{\pm}$ are the coefficients of forward moving and backward moving electrons, respectively. In this notation, $q^{\pm}_x$ is given by $q_{\pm}\cos(\theta_q^{\pm})$  with $q_{\pm}=\frac{\varepsilon_k -V_0}{s^{\prime}\hbar\upsilon_F(1\pm \Delta_0)} $, $\theta_q^{\pm}  = \sin^{-1}(k_y/q_{\pm})$ and $s^{\prime}=$ sign($\varepsilon_k -V_0$).

\medskip
\par
From a physical point of view, we should further introduce a set of boundary conditions for our model system.  Since the relationships between the four components $\psi_A(x)$, $\psi_B(x)$, $\psi_C(x)$ and $\psi_D(x)$ of a wave function  $|\Psi> =(\psi_A \ \psi_B\  \psi_C\ \psi_D)^T$ have already been built into the  eigenvalue equation, we use only the  continuities for wave functions on both sides of the boundaries at $x=0$ and $x=d$. The boundary condition at $x=0$ are
\begin{equation}
\Psi_Y^{(i)}\big|_{x=0} +\Psi_Y^{(r_+)}\big|_{x=0} + \Psi_Y^{(r_-)}\big|_{x=0} = \Psi_Y^{(\gamma_+)}\big|_{x=0} +\Psi_Y^{(\gamma_-)}\big|_{x=0} +\Psi_Y^{(\eta_+)}\big|_{x=0} +\Psi_Y^{(\eta_-)}\big|_{x=0}
\end{equation}
and  at $x=d$
\begin{equation}
\Psi_Y^{(\gamma_+)}\big|_{x=d} +\Psi_Y^{(\gamma_-)}\big|_{x=d} +\Psi_Y^{(\eta_+)}\big|_{x=d} +\Psi_Y^{(\eta_-)}\big|_{x=d} =\Psi_Y^{(t_+)}\big|_{x=d} + \Psi_Y^{(t_-)}\big|_{x=d} .
\end{equation}

From these continuity equations,  we arrive at the following relations for  the unknowns $(r_+, r_-, a_+, a_-, t_+, t_-)$, i.e.,

\begin{eqnarray}
e^{-i\theta_{k^{+}}} - r_+ e^{i\theta_{k^{+}}} - r_- e^{i \theta_{k^{-}}} = \gamma_+ e^{-i \theta_{q^{+}}} + \gamma_- e^{-i \theta_{q^{-}}} - \eta_+ e^{i \theta_{q^{+}}} - \eta_- e^{i \theta_{q^{-}}}\nonumber\\
s + s r_+ +s r_- =  s^{\prime} \gamma_+  + s^{\prime} \gamma_- + s^{\prime}\eta_+ + s^{\prime} \eta_- \nonumber\\
s + s r_+ - s r_- =  s^{\prime} \gamma_+  - s^{\prime} \gamma_- + s^{\prime}\eta_+ - s^{\prime} \eta_- \nonumber\\
e^{i\theta_{k^{+}}} - r_+ e^{- i\theta_{k^{+}}} + r_- e^{-i \theta_{k^{-}}} = \gamma_+ e^{i \theta_{q^{+}}} - \gamma_- e^{i \theta_{q^{-}}} - \eta_+ e^{-i \theta_{q^{+}}} + \eta_- e^{-i \theta_{q^{-}}}\nonumber\\
\gamma_+ e^{-i \theta_{q^{+}} + i q^{+}_x d} + \gamma_- e^{-i \theta_{q^{-}} + i q^{-}_x d} - \eta_+ e^{i \theta_{q^{+}} - i q^{+}_x d} - \eta_- e^{i \theta_{q^{-}} - i q^{-}_x d} = t_+ e^{-i \theta_{k^{+}} + i k^{+}_x d} + t_- e^{-i \theta_{k^{-}} + i k^{-}_x d} \nonumber\\
\gamma_+ s^{\prime} e^{ i q^{+}_x d} + \gamma_- s^{\prime} e^{i q^{-}_x d} + \eta_+ s^{\prime} e^{ - i q^{+}_x d} + \eta_- s^{\prime} e^{ - i q^{-}_x d} = t_+ s e^{ i k^{+}_x d} + t_-  s e^{ i k^{-}_x d} \nonumber\\
\gamma_+ s^{\prime} e^{ i q^{+}_x d} - \gamma_- s^{\prime} e^{i q^{-}_x d} + \eta_+ s^{\prime} e^{ - i q^{+}_x d} - \eta_- s^{\prime} e^{ - i q^{-}_x d} = t_+ s e^{ i k^{+}_x d} - t_-  s e^{ i k^{-}_x d} \nonumber\\
\gamma_+ e^{i \theta_{q^{+}} + i q^{+}_x d} - \gamma_- e^{i \theta_{q^{-} }+ i q^{-}_x d} - \eta_+ e^{-i \theta_{q^{-}} - i q^{+}_x d} + \eta_- e^{-i \theta_{q^{-}} - i q^{-}_x d} = t_+ e^{i \theta_{k^{+}} + i k^{+}_x d} - t_- e^{i \theta_{k^{-}} + i k^{-}_x d}.
\label{eq_boundry}
\end{eqnarray}
\medskip
\par

Since the analytical solutions of these equations  are very lengthy and time consuming we employed the numerical method explained in appendix A.  Using these results for the $t_+, t_-, r_+$ and $r_-$, one can calculate  the probability of transmission and reflection through each valley  as follows,

\begin{equation}
{\cal T}_{++}
= \frac{\left| J_x^{(t_+)} \right|  }{ \left| J_x^{(i)}  \right| } =\left|t_+\right|^2,
 \label{T_pp}
 \end{equation}
 
 \begin{equation}
{\cal T}_{+-}
= \frac{\left| J_x^{(t_-)} \right|  }{ \left| J_x^{(i)}  \right| } = \left|t_-\right|^2 \frac{(1-\Delta_0) \cos\theta_{k^{-}}}{(1+\Delta_0)\cos\theta_{k^{+}}} ,
 \label{T_pm}
 \end{equation}

\begin{equation}
{\cal R}_{++}
= \frac{\left| J_x^{(r_+)} \right|  }{ \left| J_x^{(i)}  \right| } =\left|r_+\right|^2
 \label{R_pp}
 \end{equation}
 
 \begin{equation}
{\cal R}_{+-}
= \frac{\left| J_x^{(r_-)} \right|  }{ \left| J_x^{(i)}  \right| } = \left|r_-\right|^2 \frac{(1-\Delta_0) \cos\theta_{k^{-}}}{(1+\Delta_0)\cos\theta_{k^{+}}} 
 \label{R_pm}
 \end{equation}
 so that the total transmission ${\cal T} = {\cal T}_{++} + {\cal T}_{+-}$ and the total reflection ${\cal R} ={\cal R}_{++} + {\cal R}_{+-} $.
\medskip
\par

\begin{figure}
\centering
\includegraphics[width=1\textwidth]{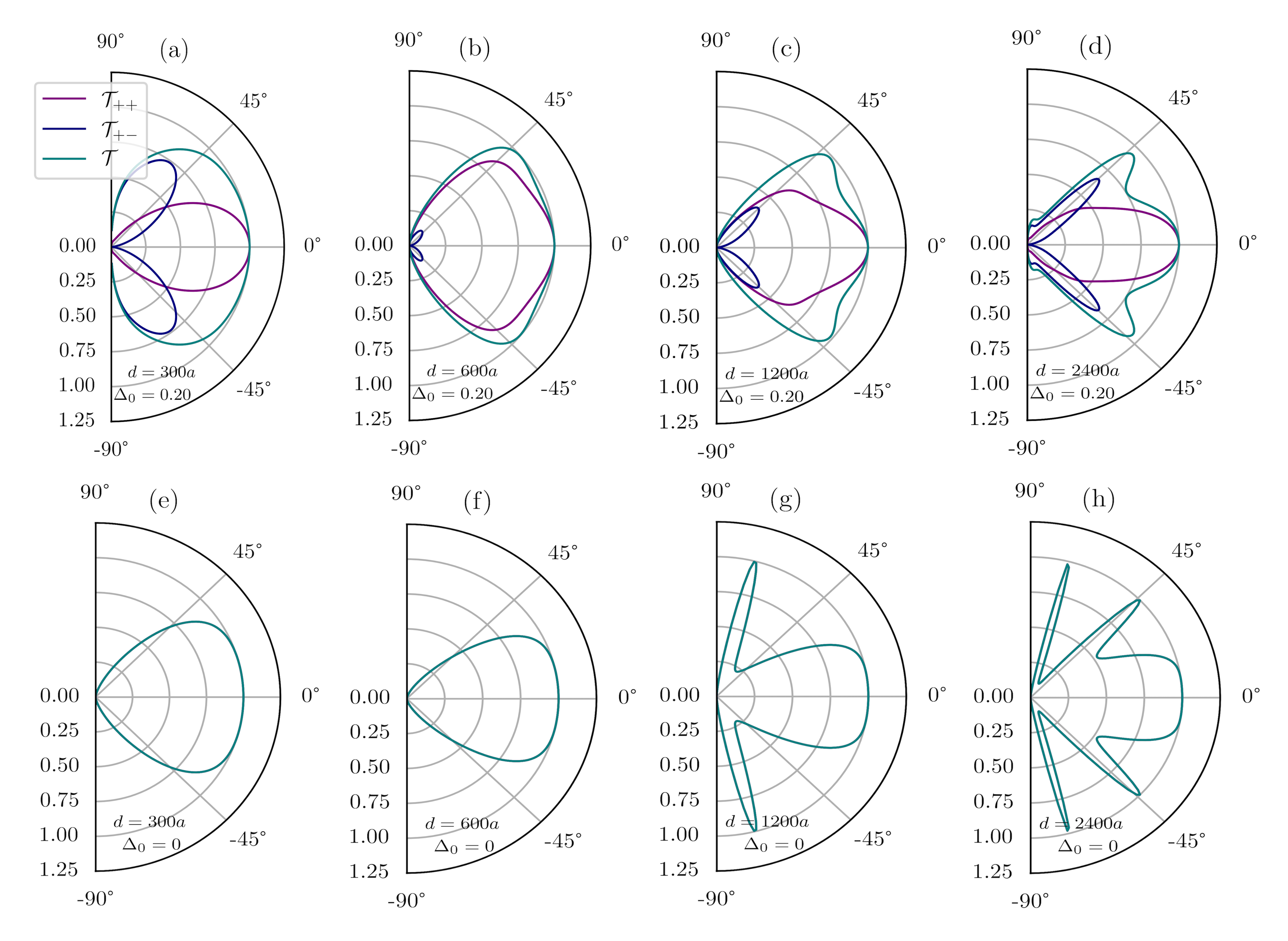}
\caption{(Color online) Valley resolved angular transmission  of monolayer  Kek$-$Y graphene  through a potential barrier of uniform height. Polar plots (a), (b), (c) and (d) in the upper panel are for barrier width $d$ equal $300a, 600a, 1200a$ and $2400a$, respectively, with Kekul\'e parameter $\Delta_0 = 0.2$.  Plots  (e), (f), (g) and (h)  in the lower panel are for undistorted graphene with  $\Delta_0 = 0$ with corresponding barrier width, in the absence of irradiation.   Here, $a = 2.46 \AA$  is the lattice constant of graphene. The energy of the incident particle is $\varepsilon_Y = 0.028$eV and the potential  height  is $V_0 = 3 \varepsilon_Y $. The notation ${\cal T}_{++}$ represents the intravalley transmission from the $\tau_+$ valley to $\tau_+$ valley while ${\cal T}_{+-}$ represents the intervalley transmission from the $\tau_+$ valley to the $\tau_{-}$ valley. The total transmission  is ${\cal T} ={\cal T}_{++}+{\cal T}_{+-}$. For undistorted graphene in the lower panel, the transmission is through only one valley. Klein tunneling at normal incidence is preserved in both distorted and undistorted graphene. }
\label{FIG:5}
\end{figure}

\medskip
\par

Let us now analyze some specific situations we encountered.
\medskip
\par
\begin{enumerate}[(i)]
\item When $\Delta_0 \neq 0$ and $1$, the two valleys at the edge of the hexagonal Brillouin  zone in monolayer gapless graphene come to the center of the Brillouin zone and get folded at the  $\Gamma$ point.  This means that  for Kek$-$Y distorted graphene, the distance between valleys is zero in reciprocal space  since one Dirac cone gets buried inside the other and hence even with the same lattice scale as pristine graphene, the potential can vary sharply causing the particle to scatter from one valley to another. Therefore, the electron from one valley has finite probability to appear as an another valley electron when it passes through the potential barrier which we refer to as intervally transmission. It is such that the total probability of transmission is the sum of the probability as an one valley as well as another valley electron. This type of valley resolved scattering opens a new topological platform for use as a valley degree of freedom. However, in pristine graphene on the lattice scale, the intervalley scattering is not possible as the separation between two valleys in reciprocal space is equal to the reciprocal of the lattice constant. Under this condition, the potential varies smoothly and valleys are independent.  Hence, intervalley scattering is forbidden. The numerical results for the transmission probability at different angles and chosen width of the barrier for Kek$-$Y distorted graphene at $\Delta_0 =0.2$ are presented in  Fig.\ \ref{FIG:5}. In the displayed figure, the transmission probability is $1$ around  normal incidence  regardless of width of the barrier for potential three times higher than that  of incident energy  $\varepsilon_Y$.  This perfect transmission even for the barrier higher than that of the energy of incident beam is known as  Klein tunneling \cite{Geim}.  However,  Allain  in Ref.\ [\onlinecite{allain}], has argued that perfect transmission at normal incidence is not a genuine Klein tunnelling as it does not incorporate any classically forbidden region and neither even any evanescent or quickly decaying waves are present. It is due to the presence of the valance band state and conduction band state at the boundary of the potential where they matched, and also due to  the pseudospin  conservation or due to the pseudo-spin flip at two sublattices where the waves are continuous to move forward.  This is discussed as the chiral behavior of graphene that prevents backscattering and gives perfect transmission for the normally incident beam. 
\medskip
\par
As we see from Fig.\ref{FIG:5}, the transmission does not only depend on the angle of incidence, it also is a function of the incident particle energy, height and width of the barrier and also the Kekul\'e parameter $\Delta_0$. For example,  the region for perfect transmission for $d= 300a$ is about $\theta_{k^{+}}\le \pi/6$ and is reduced when the barrier width is increased. It is true only for normal incidence for a potential step with infinite barrier width. It is also seen that the intervalley scattering does not exist at around normal incidence and there is perfect reflection at $\theta_{k^{+}} =\pi/2$. 
 
 \item When $\Delta_0 =0$, the energy dispersion comes out to be that of pristine graphene. In this case, the numerical results for  the transmission and reflection probability reproduce those for gapless graphene  in   Ref. [\onlinecite{dipendra}] .  Figure\  \ref{FIG:5} shows  that in the lower panel for $\Delta_0$ equal to zero, the transmission  is only due to  intravalley transmission. For gapless monolayer graphene for normal incidence i.e., $\theta_{\bf k^{+}} = 0$,   $|{\cal T}| =1$ irrespective of the kinetic energy of the incident beam. For wider potential barrier, perfect transmission is observed at some points besides normal incidence  or it oscillates between 1 and 0 similar to the sine and cosine functions which is due to resonance scattering. Such resonance tunneling is suppressed when the intervalley transmission is taken into account for $\Delta_0 \neq0$.  In contrast to perfect transmission through monolayer graphene, bilayer graphene shows perfect reflection at normal incidence\cite{Geim}. Although both the negative energy states and positive energy states are present at the boundary, chirality does not allow the transmitted wave to travel in the forward direction. Since the Berry phase of bilayer graphene is $2\pi$ instead of $\pi$ the pseudospin is not conserved between two sublattices, creating the wave quickly, vanishes in the potential region, enhancing the backscattering to the incident region. 
 
  \item When $\Delta_0 =1$, the outer valley for $\tau=-1$, becomes dispersionless. At around normal incidence, all the electrons from outer flat band couples to the inner dispersive band and hence  crosses the junction  giving rise to super-Klein tunneling \cite{review}. This extreme condition has not been considered for calculation which is beyond the scope of this manuscript.  
 \medskip
\par

We should mention that in the above calculation  only the  longitudinal momentum $k_x$ is changes, keeping the transverse momentum $k_y$ conserved assuming translational symmetry along the $y$ direction. Therefore  the observed transmission is  for the longitudinal transmission. Besides the translational symmetry, the material also preserves inversion symmetry. We     note that as we are interested in seeing Klein tunneling, we have considered the potential much higher than that of the energy of incident particle which is equal to $0.028$ eV in our numerical results.   

\end{enumerate}

\medskip
\par

\subsubsection{Electrical conductivity using Landauer-Buttiker formalism}

In quantum mechanics, the electrical conductance can be calculated by using the Landauer formula for ballistic transport \cite{zitt} at 0 K, which necessitates a calculation of the transmission probability. It is given by 

\begin{equation} 
{\cal G}= \frac{2e^2}{h} {\cal  T}
\end{equation}
where ${\cal  T}$ is the electron transmission probability through all possible modes. For specific valley and spin orientation, it can be expressed as

\begin{equation} 
{\cal G}= {\cal G}_0 \int_{-\pi/2}^{\pi/2} d\theta_k \cos\theta_k {\cal  T}(\theta_k)
\label{conductance}
\end{equation}
where ${\cal G}_0 =\frac{2e^2}{h} $ is the conductance quantum and  ${\cal  T}(\theta_k) = \sum_{\tau,\tau^{\prime}} {\cal  T}_{\tau,\tau^{\prime}}(\theta_k)$, for $\tau, \tau^{\prime} = \pm$.

\medskip
\par
\begin{figure}
\centering
\includegraphics[width=0.45\textwidth]{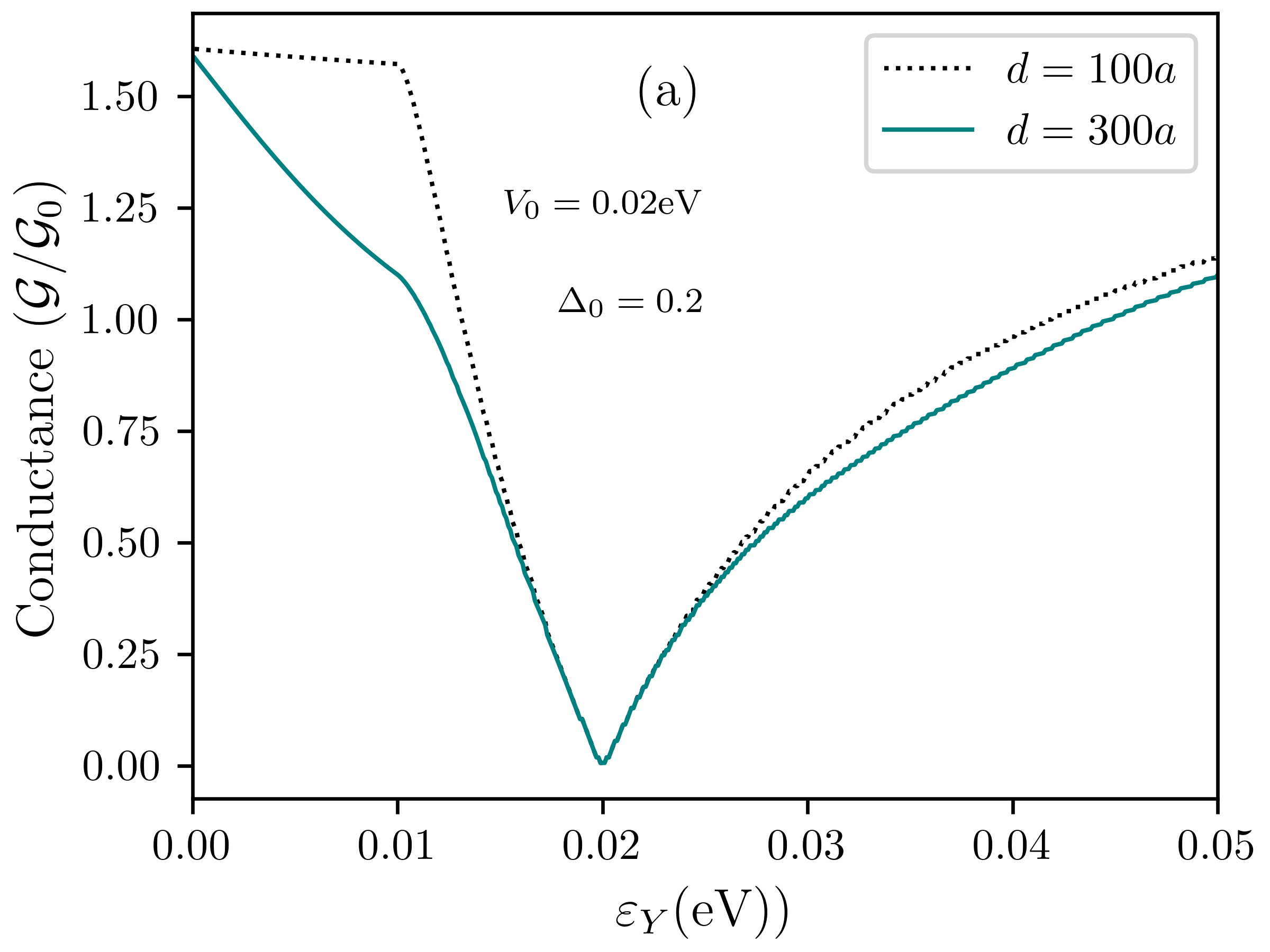}
\includegraphics[width=0.45\textwidth]{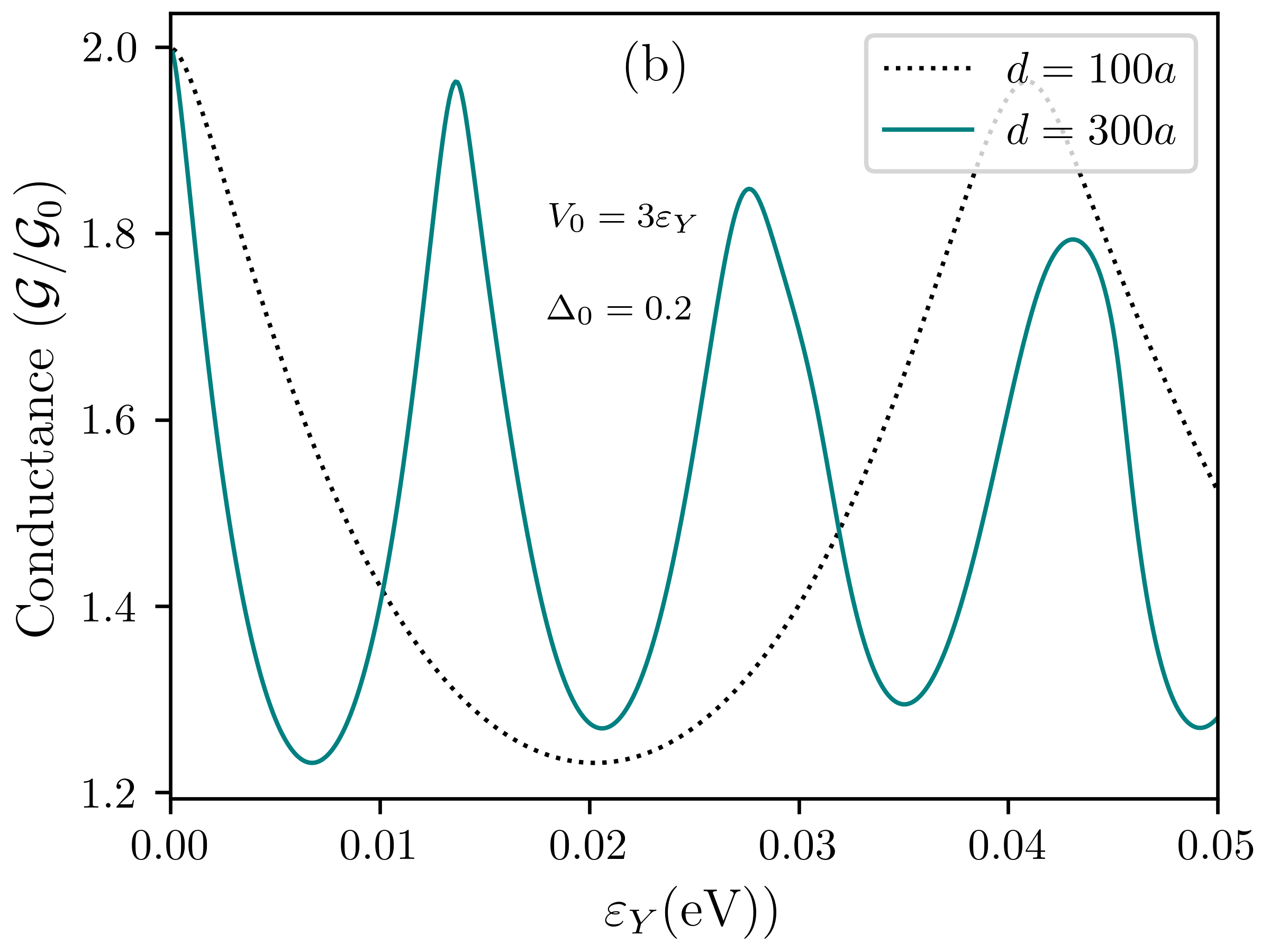}
\caption{(Color online)  Quantum conductance of ballistic transport for Kek$-$Y distorted graphene across a potential barrier of width $d$.  The coefficient  of conductance ${\cal G}/ {\cal G}_0 $ is plotted as a function of the kinetic energy of an incident electron for chosen potential height $V_0 = 0.02$eV in (a) and for barrier height equal to $3$ times the incident energy in (b). }
\label{FIG:6}
\end{figure}

In Fig.\  \, (\ref{FIG:6}), we present numerical results for the electrical conductance  through  a potential barrier in Kek-Y graphene, in the absence of irradiationas as a function of kinetic energy for an electron in the presence of  a barrier of height larger than the kinetic energy of the incident electron. As the energy is increased, the conductance exhibits periodic oscillations and we observe a dip in the conductance when the electron energy is close to the barrier height. This is due to the reduction in the transmission probability in that regime. 
\medskip
\par

\medskip
\par
\subsection{Effect on tunneling due to circularly  Polarized Irradiation}
 
 We now focus our attention on the effect due to irradiation on transmission and reflection of particles across a  potential barrier. With the Hamiltonian in Eq.\, (\ref{HAC}), the energy eigenvalues in Eq.\,(\ref{MEX}) and eigenvectors in  Eq.\,\eqref{wave1C}, we obtain the  probability current density $j_x$ along the longitudinal direction as 
 
 \begin{eqnarray}
j_x &=&\upsilon_f  \left[  (\psi^{*}_A \psi_B +\psi^{*}_B \psi_A +\psi^{*}_C \psi_D +\psi^{*}_D \psi_C) + \Delta_{0} (\psi^{*}_A \psi_C +\psi^{*}_C \psi_A +\psi^{*}_B \psi_D +\psi^{*}_D \psi_B)\right] \nonumber\\
&+&   2\upsilon_f \tilde{\zeta} (1+ \Delta_0^2)(\psi^{*}_A \psi_A -\psi^{*}_D \psi_D)  +   2\upsilon_f \tilde{\zeta} (1- \Delta_0^2) ( \psi^{*}_C \psi_C -\psi^{*}_B \psi_B)  \nonumber\\
\label{jx1}
\end{eqnarray}

Additionally, in the presence of irradiation, we consider an electron from the $\tau_+ $ valley, which yields the incident current
 
 \begin{eqnarray}
J_x^{(\tilde i)} &=& 2\upsilon_f  \left[ \left(\frac{\text{b}_s(\theta_{k^{+}})}{1+\text{b}_s^2(\theta_{k^{+}}) }   + \Delta_0 \frac{\text{b}_{\tau}(\theta_{k^{+}})}{ 1+\text{b}^2_{\tau}(\theta_{k^{+}}) } \right) \cos\theta_{k^{+}}   + \tilde\zeta \left( \frac{1-\text{b}^2_s(\theta_{k^{+}})}{1+\text{b}^2_s(\theta_{k^{+}})}  + \Delta^2_0 \frac{1-\text{b}^2_{\tau}(\theta_{k^{+}})}{1+\text{b}^2_{\tau}(\theta_{k^{+}})}    \right)\right]
\label{jx_i_til}
\end{eqnarray}
with the corresponding incident wave function  

\begin{equation}
\label{wave_i_c}
\Psi^{E(i)}_{Y}({\bf k};{\bf r}) = \  \frac{1}{\sqrt{(1+ \text{b}^{2}_{s}(\theta_{k^{+}}))(1+\text{b}^{2}_{+}(\theta_{k^{+}}) )}} \, \left(
\begin{array}{c}
\text{e}^{- i \theta_{k^{+}}} \\
\text{b}_s(\theta_{k^{+}})  \\
\text{b}_+(\theta_{k^{+}})  \\
\text{b}_s(\theta_{k^{+}})  \text{b}_+(\theta_{k^{+}}) \text{e}^{ i \theta_{k^+}}          
\end{array}
\right)
\frac{e^{ik_y y + i| k^+_x |x}}{\sqrt{A}}\ .  
\end{equation}

Similarly, the transmitted and reflected current with transmitted and reflected wave functions are presented in the Appendix B. Here, the parameters are defined as follows

\begin{equation}
k_{\tau} = \frac{\varepsilon^{E}_Y}{s\hbar\upsilon_F [(1+\tau\Delta_0) + 2\tilde\zeta^2 \cos^2\theta_{k^{\tau}} (1+\tau \Delta_0^3)]}
\end{equation}
with 

\begin{equation}
\varepsilon^{E}_Y=  s \left\{ \sqrt{C^2 + \hbar^2 \upsilon^2_f k^2}  + \tau  \Delta_{0}
\sqrt{C^2 \Delta^2_{0}+ \hbar^2 \upsilon^2_f k^2} \right\}
\end{equation}
and $C = \tilde\zeta^2\hbar\upsilon - 2\tilde\zeta \hbar\upsilon_Fk \cos\theta_k$. We assume that, $\zeta<<\hbar\Omega$, $\zeta<<\hbar\upsilon_Fk$ and perform a Taylor expansion from which we obtain $C= -2\tilde\zeta\hbar\upsilon_Fk\cos\theta_k$ and $\varepsilon^{E}_Y = s\hbar\upsilon_Fk [(1+\tau\Delta_0) +2\tilde\zeta^2 \cos^2\theta_k (1+\tau\Delta_0^3)]$, where $\tilde\zeta =\zeta /\hbar \Omega$ so that we neglected the higher order terms of $\tilde\zeta$. From the expression for $\varepsilon^{E}_Y$ and $k_{\tau} $, we can calculate $k^{\tau}_{x} = k_{\tau} \cos\theta_{k^{\tau}}$ and $k_{y} = k_{\tau} \sin\theta_{k^{\tau}}$,  ($\tau=+$ for an incident particle from the $\tau_+$ valley). Similarly, $\text{b}_s( \theta_{k^{\tau}}) = s(1+2\tilde\zeta^2 \cos^2\theta_{k^{\tau}} -2s\tilde\zeta\cos\theta_{k^{\tau}})$ and $\text{b}_{\tau}( \theta_{k^{\tau}}) = s\tau/ (1+2\Delta_0^2\tilde\zeta^2 \cos^2\theta_{k^{\tau}} +2s\tau\Delta_0\tilde\zeta\cos\theta_{k^{\tau}})$, where $s =$sign$(\varepsilon^{E}_Y)$.

\medskip
\par
The incident, reflected and transmitted waves are from the (I) and (III) regions in the schematic in Fig.\  \ref{FIG:4}. In the scattering region (II), the forward moving and backward moving waves with their corresponding coefficients are given in  the Appendix B.  Making use of the boundary conditions at $x=0$ and $x=d$, we solve for ${\tilde r}_+,  {\tilde r}_-, {\tilde t}_+, {\tilde t}_-$.

\begin{figure}
\centering
\includegraphics[width=1\textwidth]{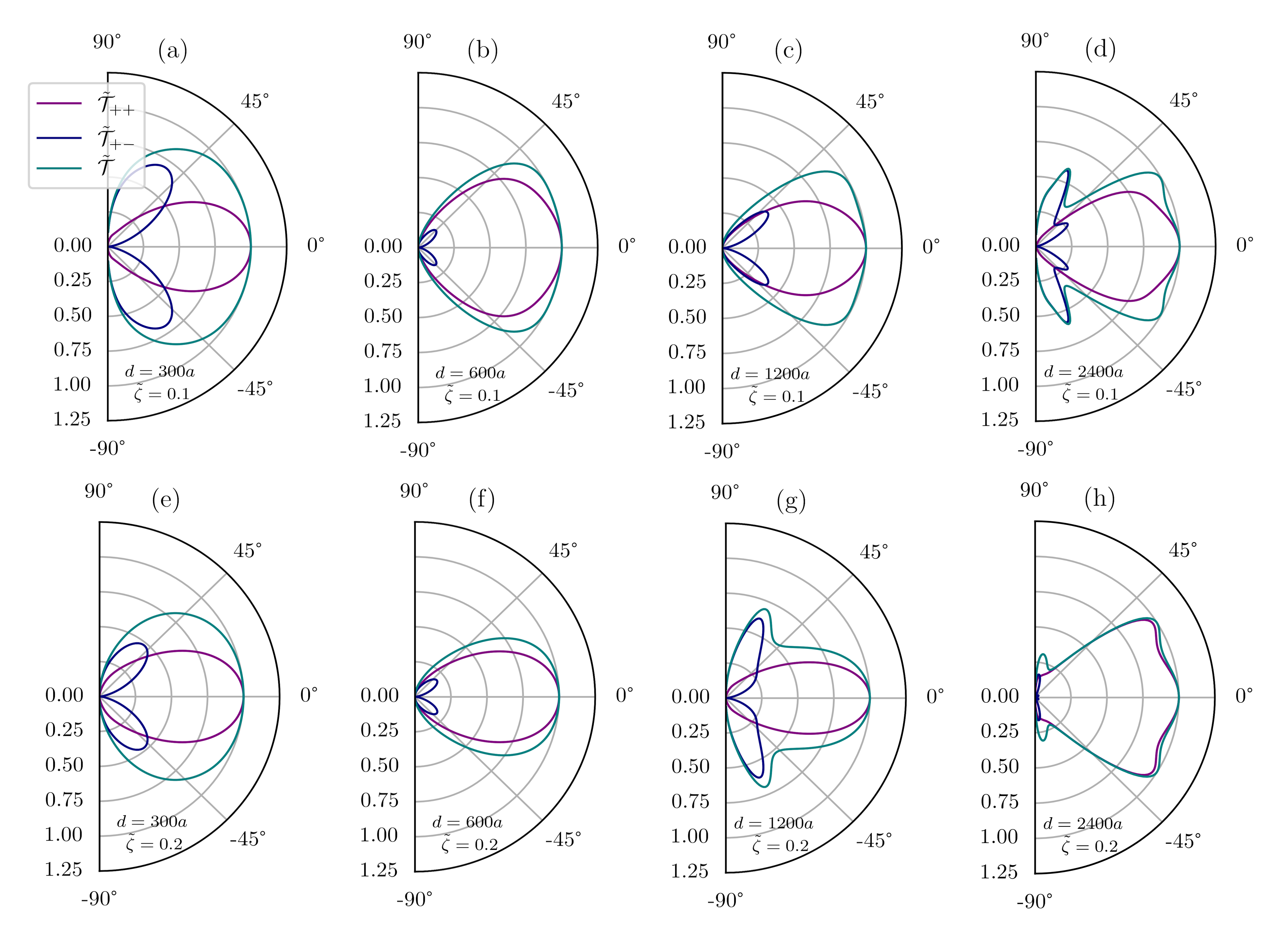}
\caption{(Color online) Valley resolved angular transmission  of monolayer  Kek$-$Y graphene  across a potential barrier of uniform height in the presence of circularly polarized light. Plots (a), (b), (c) and (d) in the upper panel are for barrier width $d$ equal to $300\,a, 600\,a, 1200\,a$ and $2400\,a$, respectively, for radiation parameter $\tilde \zeta = 0.1$ . Plots (e), (f), (g) and (h)  in the lower panel are for $\tilde \zeta = 0.2$  for corresponding barrier width.   Here, $a = 2.46 \AA$  is the lattice constant of graphene. The energy of the incident particle is $\varepsilon_Y = 0.028$ eV and the height of the potential is chosen as $V_0 = 3 \varepsilon_Y $. We have chosen the Kekul\'e parameter $\Delta_0 = 0.2$. In our notation, $\tilde{\cal T}_{++}$ represents intravalley transmission from the $\tau_+$ valley to the $\tau_+$ valley while $\tilde {\cal T}_{+-}$ representing intervalley transmission from the $\tau_+$ valley to the $\tau_-$ valley. The total transmission $\tilde {\cal T} =\tilde {\cal T}_{++}+\tilde{\cal T}_{+-}$.  Klein tunneling at normal incidence is preserved even in the presence of high-frequency radiation. }
\label{FIG:7}
\end{figure}

\medskip
\par

\begin{figure}
\centering
\includegraphics[width=1\textwidth]{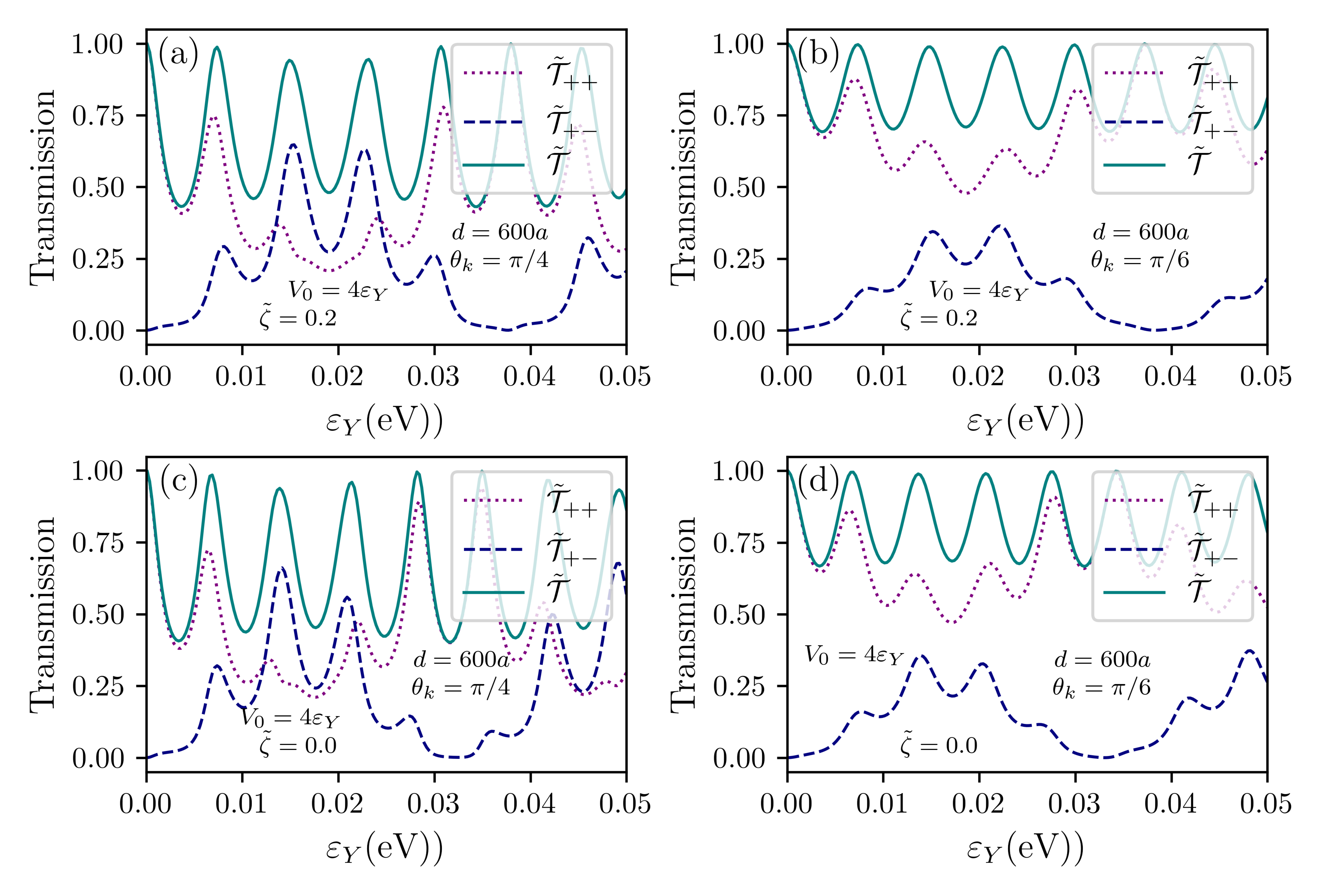}
\caption{(Color online) Valley resolved  transmission  of monolayer  Kek-Y graphene  through a potential barrier of constant height as a function of  kinetic energy of the  incident particle. Plots (a) and (b) in the upper panel are  in the presence of circularly polarized light with irradiation parameter $\tilde\zeta = 0.2$   for incident angles $\pi/4$ and $\pi/6$, respectively, whereas plots (c) and (d)  in the lower panel are for the same angle of incidence but in the absence of irradiation.  The  chosen height of the potential barrier is $V_0 = 4 \varepsilon_Y $ and the width is $d=600\,a$.   Here, $a = 2.46 \AA$  is the lattice constant of graphene. We have chosen the Kekul\'e parameter $\Delta_0 = 0.2$. The symbol $\tilde{\cal T}_{++}$ represents the intravalley transmission from the $\tau_+$ valley to the $\tau_+$ valley while $\tilde {\cal T}_{+-}$ denotes intervalley transmission from the $\tau_+$ valley to the $\tau_-$ valley.  The total transmission $\tilde {\cal T} =\tilde {\cal T}_{++}+\tilde{\cal T}_{+-}$. }
\label{FIG:8}
\end{figure} 

Now,   the probability of the transmission and reflection through each valley in the presence of irradiation in terms of these coefficients can be expressed  as follows,

\begin{eqnarray}
{\cal \tilde T}_{++}
= \frac{\left| J_x^{({\tilde t}_+)} \right|  }{ \left| J_x^{(i)}  \right| } =\left|{\tilde t}_+\right|^2,
 \label{T_pp_Ir}
 \end{eqnarray}
 
 \begin{eqnarray}
{\cal \tilde T}_{+-}
= \frac{\left| J_x^{({\tilde t}_-)} \right|  }{ \left| J_x^{(i)}  \right| } =\left|{\tilde t}_-\right|^2 \alpha
 \label{T_pm_Ir}
 \end{eqnarray}

 \begin{eqnarray}
{\cal \tilde R}_{++}
= \frac{\left| J_x^{({\tilde r}_+)} \right|  }{ \left| J_x^{(i)}  \right| }= \left|{\tilde r}_+\right|^2 \lambda
 \label{R_pp_Ir}
 \end{eqnarray}   
  and 
 
  \begin{eqnarray}
{\cal \tilde R}_{+-}
= \frac{\left| J_x^{({\tilde r}_-)} \right|  }{ \left| J_x^{(i)}  \right| } = \left|{\tilde r}_-\right|^2 \kappa
 \label{R_pm_Ir}
 \end{eqnarray}
 so that the total transmission ${\cal \tilde T} = {\cal \tilde T}_{++} + {\cal \tilde T}_{+-}$ as well as the total reflection ${\cal \tilde R} ={\cal \tilde R}_{++} + {\cal \tilde R}_{+-} $. Here, $\alpha, \Lambda$ and $\kappa$ are given in the Aappendix B.

\medskip
\par
 The application of circularly polarized irradiation on Kek$-$Y graphene opens up a small gap between the valance and conduction bands near the Dirac point.   It also produces a small energy difference near the minima of  the two folded valleys but it does not affect the chirality, the same as in the absence of irradiation. The bands become anisotropic along the longitudinal momentum direction  for $\theta_p =0$ and hence the inversion symmetry is broken. In a recent paper, Wang\cite{juan} has shown that addition of site energy modification in Y-shaped Kekul\'e pattern breaks the inversion symmetry and which also enhances Klein tunneling.  Our numerical results showing  the  effect due  to circularly polarized  irradiation on valley resolved transmission of  an electron across the potential barrier in Kek$-$Y graphene are presented in Fig.\, \ref{FIG:7}.  We note that the irradiation is not so strong  since we have considered the illumination energy to be about $0.1$ eV.  All other parameters employed to calculate the results are revealed in the figure caption. The results shown in the upper panel of the figure were obtained in the presence of radiation with parameter ${\tilde \zeta} = 0.1$. This means that the frequency of the impinging electromagnetic wave is about  $100$ THz and the intensity of the field is about 3000 V/m. In this case, the transmission is not significantly affected for the smaller barrier width. However, as the barrier  width is increased,  the transmission is reduced for incident angle other than normal incidence. Perfect transmission for normal incidence is still preserved even in the presence of an EM field. Although the inversion symmetry is broken due to irradiation it did not enhances the transmission.  The reduction of the transmission probability seems  to be a consequence of intervalley scattering. When the eigenstates are dressed ,they are less likely to change their velocity in the scattering region. The range of the angle of incidence for perfect transmission  is further reduced when the radiation intensity is made stronger. Which is seen in the plots in the lower panel of Fig.\, \ref{FIG:7} which were computed for $\zeta = 0.2$.  It suggests that illuminating the material with electromagnetic irradiation does not  enhance the  conditions for Klein tunneling in Kek$-$Y graphene although irradiation has other significant benefits in 2D materials.  
 
 \medskip
\par

Since the gap opening  is due to irradiation, one should know what would be the effect on the transmission when we  change the energy of the incident particle around the gap. This is one of the interesting events we would like to address. For this, we have shown the transmission as a function of incident energy in Fig.\,\ref{FIG:8} for two chosen incident angles. For the given parameters in the figure, a band gap of approximately $5.6$ meV is opened and the gap between two valleys is around $2.5$ meV. In the plots (a) and (b) of Fig. \ref{FIG:8},  the radiation parameter is chosen  as $0.2$ and the potential is four times that of   the incident energy. The total transmission is oscillatory  between $0.5$ and $1.0$  for an incident angle $\pi/4$ and between $0.75$ and $1.0$ for incident angle $\pi/6$ as a function of energy. When the energy of an incident particle is very small in the vicinity of the gap,  and the potential is very high, the transmission is totally due to intravalley scattering. However, when the incident energy is increased and becomes larger than the gap, the intervalley scattering becomes dominant. For comparison, the results in the absence of irradiation are plotted in (c) and (d) in  the lower panel.   All these results show that the transmission depends on all the factors such as the energy of the incident particle, the potential width and height, irradiation parameter and angle of incidence.  If we carefully choose these quantities for an experiment,   we can tune the Klein tunneling and hence the electrical conductivity of the material which we will discuss in another section.

\section{Exciton-assisted tunneling in gapped Kek$-$Y graphene}
\label{sec5}

The quantum tunneling of a bound electron-hole pair across a potential barrier is governed by its mass in the center of mass  frame of reference as  well as the binding energy of the pair. Since the binding energy of the exciton  is quite large and due to its finite effective mass,  the tunneling of excitons through an atomic scale barrier is significant in a 2D material. Here, we use the general formalism presented by  Kavka \cite{kav1}      for quantum mechanical tunneling and reflection of an exciton in a single bound state incident upon a potential barrier   to demonstrate the multichanel transport of excitons. We will also present the transmission and reflection probability dependent on the irradiation parameter which plays a role  in opening an energy bandgap for the system.
\medskip
\par
Let us consider an electron of mass $m_e$,  with  position coordinate $x_e$, in the conduction band  and a hole of mass $m_h$, with coordinate $x_h$,  in the valance band  of Kek$-$Y graphene under circularly polarized light. These two particles are     are bound by the attractive Coulomb potential $V(r)= -e^2/\epsilon_s |x|$ where $\epsilon_s\equiv 4\pi\epsilon_0\epsilon_b$ with $\epsilon_b$ denoting the background dielectric constant. The electron experiences an external potential barrier $V_e(x_e)$ and the hole is subjected to $V_h(x_h)$.  The Hamiltonian for this system is given by, 

\begin{equation}
{\bf H} = -\frac{\hbar^2}{2M} \frac{\partial^2}{\partial X^2} -\frac{\hbar^2}{2\mu} \frac{\partial^2}{\partial x^2}  -i \hbar \frac{3\upsilon_{Fx}}{2} \frac{\partial}{\partial X}+ V_e(X- \frac{\mu}{m_e} x) + V_h(X + \frac{\mu}{m_h} x)  - \frac{e^2}{\epsilon_s |x|}\  ,
\label{exciton_hamil}
\end{equation}
where $M = m_e + m_h$ and $\mu =m_e m_h/M$  are center of mass and relative mass of the electron-hole pair and $X = \frac{m_e x_e + m_h x_h}{m_e+ m_h}$ and $x = x_e - x_h$ are their corresponding coordinates.  The third term proportional to $\upsilon_{Fx}$ is due to the linear term in the Taylor expansion of  the energy eigenvalue in Eq. \ (\ref{MEX}). Also, $\upsilon_{Fx} = 2\upsilon_F {\tilde\zeta} (1+\tau \Delta_0^2)$ is the group velocity and $m_{e,h}= \frac{\tilde\zeta^2 \hbar\Omega}{\upsilon_F^2 (1+\tau)}$ is the effective mass of the electron or hole. Since for the outer valley, $\tau=-1$, the mass diverges. This signifies  that the zero curvature of the electron subband near the Dirac point for dressed states is due to circularly polarized irradiation. The electron or hole in this valley does not contribute to the transmission or electrical conductivity. Therefore, we will confine our numerical calculations for electron or hole tunneling  from the inner valley with $\tau=+1$ . 
We solved for the eigenvalue and eigenfunction for the exciton by using the method of separation of variables.  The solutions of the one dimensional hydrogen atom problem were discussed in our paper. \cite{sita} These are , 

\begin{equation}
E_p = - \frac{\mu (e^2/\epsilon_s)^2}{2\hbar^2 (p-1/2)^2} = -\frac{\hbar^2}{2\mu a_{e}^{2} (p-1/2)^2}
\label{eign_exciton}
\end{equation}
and 
\begin{equation}
\phi_p(z) = N_p z e^{-|z|/2} L_{p+1}^1 (|z|)
\label{eign_fun_exciton}
\end{equation}
 where $p = 0, 1, 2, \cdots$, $z = 2x/[(p+1) a_e]$, $N_p= 1/[2(p+2)]$ is a normalization constant and $a_e = \hbar^2 \epsilon_s/e^2 \mu$ is the exciton Bohr radius. With these values for the eigenfunctions and eigenvalues, the Schr\"odinger-Dirac equation becomes
 
 \begin{equation}
 {\bf H}\psi(x, X) = \varepsilon_Y \psi(x,X)
 \end{equation}
 such that $\psi(x,X) = \phi_p (x) \xi_p(X)$ and $\varepsilon_Y = E_p + E^{\prime}_p$.

 \begin{equation}
\left( \frac{d^2}{dX^2} + i\hbar \frac{6m_{e,h}\upsilon_{Fx}}{\hbar^2} \frac{d}{dX}   + K^2_m\right) \xi_m(X)  + \sum_p Z_{mp}(X) \xi_p(X) = 0\  ,
 \label{schr_exciton}
 \end{equation}
where, $Z_{mp}(X) = \frac{4m_{e,h}}{\hbar^2} \int_{-\infty}^{\infty} dx\  \phi^{\ast}_m (x) \phi_p (x)[V_e (X-x/2) + V_h (X+x/2)]$ is the effective potential for the center of mass and $K_m^2 = \frac{4m_{e,h}}{\hbar^2}(\varepsilon_Y - E_m) $ is the center of mass wave number for an exciton in state $m$. The effective potential  $Z_{mp}(X)$ can be thought of as the potential barrier that the center of mass encounters while incident in the state $p$  and reflected or transmitted in the state $m$.  Equation (\ref{schr_exciton}) can be rewritten in the form as
 
 \begin{equation}
\left( \frac{d^2}{dX^2}   + K^{\prime 2}_m\right) \xi^{\prime}_m(X)  + \sum_p Z_{mp}(X) \xi^{\prime}_p(X) = 0\  ,
 \label{schr_exciton_modified}
 \end{equation}
where $K^{\prime 2}_m = K_m^2 -\left( \frac{3m_{e,h}\upsilon_{Fx}}{\hbar}\right)^2 $, $ \xi_m(X)= e^{-i \frac{3m_{e,h}\upsilon_{Fx}}{\hbar} X} \xi^{\prime}_m(X)$ and  $ \xi_p(X)= e^{i \frac{3m_{e,h}\upsilon_{Fx}}{\hbar} X} \xi^{\prime}_p(X)$.

\medskip
\par
If a plane wave from left of the barrier is incident in the $n^{th}$ state of the exciton, then we can find the formal solution of Eq.\ (\ref{schr_exciton_modified}) is 

\begin{equation}
\xi^{\prime}_{mn}(X) = e^{i K_{m}^{\prime } X} \delta_{mn}+ \frac{1}{2i K_{m}^{\prime}} \sum_{p=0}^{\infty} \int_{-\infty}^{\infty} e^{iK_{m}^{\prime}|X-X^{\prime}|} Z_{mp}(X^{\prime})\xi_{pn}^{\prime}(X^{\prime}) dX^{\prime} \  .
\label{sol_modified_exciton}
\end{equation}
Since $ \xi_m(X)= e^{-i \frac{3m_{e,h}\upsilon_{Fx}}{\hbar} X} \xi^{\prime}_m(X)$ and  $ \xi_p(X)= e^{i \frac{3m_{e,h}\upsilon_{Fx}}{\hbar} X} \xi^{\prime}_p(X)$, the solution of Eq.\ (\ref{schr_exciton}) which is employed to calculate the reflection and transmission coefficient and is given by 

\begin{equation}
\xi_{mn}(X) = e^{i\left(K_{m}^{\prime} -\frac{3m_{e,h}\upsilon_{Fx}}{\hbar} \right)X}\delta_{mn} +  \frac{1}{2i K_{m}^{\prime}} \sum_{p=0}^{\infty} \int_{-\infty}^{\infty} dX^{\prime} \     e^{i\left(K_{m}^{\prime}-\frac{3m_{e,h}\upsilon_{Fx}}{\hbar}\right)|X-X^{\prime}|} Z_{mp}(X^{\prime})\xi_{pn}(X^{\prime}) \  ,
\label{sol_schr_exciton}
\end{equation} 
which we use to define the reflection and transmission amplitudes, i.e.,

\begin{equation}
R_{mn}= \frac{1}{2i K_{m}^{\prime}} \sum_{p=0}^{\infty} \int_{-\infty}^{\infty} dX^{\prime}\  e^{i\left(K_{m}^{\prime}-\frac{3m_{e,h}\upsilon_{Fx}}{\hbar}\right)X^{\prime}} Z_{mp}(X^{\prime})\xi_{pn}(X^{\prime}) 
\label{ref_amplitude}
\end{equation}
and 
\begin{equation}
T_{mn}= \delta_{mn} + \frac{1}{2i K_{m}^{\prime}} \sum_{p=0}^{\infty} \int_{-\infty}^{\infty} dX^{\prime}\  e^{-i\left(K_{m}^{\prime}-\frac{3m_{e,h}\upsilon_{Fx}}{\hbar}\right)X^{\prime}} Z_{mp}(X^{\prime})\xi_{pn}(X^{\prime}) \  .
\label{trns_amplitude}
\end{equation}
Next, we followed the procedure developed by Razavy\cite{razavy}, and we obtained coupled differential equations for the variable reflection and transmission coefficients as,

\begin{eqnarray}
\frac{d R_{mn}(y)}{dy} &= &-\sum_{j=0}^{\infty} \frac{1}{2i K_{j}^{\prime}} \left\{e^{i\left(K_{j}^{\prime}-\frac{3m_{e,h}\upsilon_{Fx}}{\hbar}\right)y} \delta_{mj} + R_{mj}(y)e^{-i\left(K_{j}^{\prime}-\frac{3m_{e,h}\upsilon_{Fx}}{\hbar}\right)y} \right\} \nonumber\\
&\times& \sum_{p=0}^{\infty}Z_{jp}(y) \left\{e^{i\left(K_{p}^{\prime}-\frac{3m_{e,h}\upsilon_{Fx}}{\hbar}\right)y} \delta_{pn} + R_{pn}(y)e^{-i\left(K_{p}^{\prime}-\frac{3m_{e,h}\upsilon_{Fx}}{\hbar}\right)y} \right\} 
\label{diff_ref}
\end{eqnarray}
and 

\begin{eqnarray}
\frac{d T_{mn}(y)}{dy} &= &-\sum_{j=0}^{\infty} \frac{1}{2i K_{j}^{\prime}} \left\{e^{i\left(K_{j}^{\prime}-\frac{3m_{e,h}\upsilon_{Fx}}{\hbar}\right)y}   T_{mj}(y) \right\} \nonumber\\
&\times& \sum_{p=0}^{\infty}Z_{jp}(y) \left\{e^{i\left(K_{p}^{\prime}-\frac{3m_{e,h}\upsilon_{Fx}}{\hbar}\right)y} \delta_{pn} + R_{pn}(y)e^{-i\left(K_{p}^{\prime}-\frac{3m_{e,h}\upsilon_{Fx}}{\hbar}\right)y} \right\} 
\label{diff_trans}
\end{eqnarray}
with the boundary conditions

\begin{eqnarray} 
R_{mn}(y\to \infty) = 0 \,\,\,\,\,\,\,\,\, R_{mn}(y\to -\infty) = R_{mn}\nonumber\\
T_{mn}(y\to \infty) = \delta_{mn} \,\,\,\,\,\,\,\,\, R_{mn}(y\to -\infty) = T_{mn}.
\label{boundry_cond}
\end{eqnarray}

By solving these equations for $R_{mn}$ and $T_{mn}$  we will calculate the total probability of transmission and reflection of an exciton incident on the state $n$ , which is given by the expressions,

\begin{eqnarray}
R = \sum_{m} \frac{K_{m}^{\prime}}{K_{n}^{\prime}} |R_{mn}|^2\nonumber\\
T = \sum_{m} \frac{K_{m}^{\prime}}{K_{n}^{\prime}} |T_{mn}|^2\   .
\label{prob_RT}
\end{eqnarray}
In this notation,$K_{m}^{\prime} = \left\{\frac{4 m_{e,h} (\varepsilon_Y - E_m)}{\hbar^2}  - \left(\frac{3 m_{e,h} \upsilon_{Fx}}{\hbar}\right)^2\right\}^{1/2}$ with $m_{e,h}= \frac{\tilde\zeta^2 \hbar\Omega}{\upsilon_F^2 (1+\tau)}$  and $E_m =  -\frac{\hbar^2}{m_{e,h} a_{e}^{2} (m-1/2)^2}$\, .

\medskip
\par
We solved the two coupled differential equations \ (\ref{diff_ref}) and (\ref{diff_trans}) numerically for two different potential profiles.  For the numerical solution, we introduced the dimensionless parameters $\tilde{m}_{e,h}= \frac{m_{e,h}}{m_0}$, $\tilde{X}= \frac{X}{a_0}, \tilde{x}= \frac{x}{a_0}, \tilde{a}_e= \frac{a_e}{a_0}, \tilde{Z}_{mp}  =a_0^2 Z_{mp}, \tilde{K}_m= a_0 K_m , \tilde{\upsilon}_{Fx} = \frac{\upsilon_{Fx}}{\upsilon_F}$ and $\tilde{E}_{m}= \frac{E_m 2 m_0 a_0^2}{\hbar^2}$ where $ \frac{\hbar^2}{ 2m_0 a_0^2} =E_{H}= 13.6$ eV is the binding energy of hydrogen atom with the rest mass of electon $m_0$ and the electron Bohr radius $a_0$. These parameters  themselves are   functions of  the semiconducting material where the exciton is formed. For irradiated gapped Kek$-$Y graphene with parameter $\tilde{\zeta} =0.1$ at high frequency ($\hbar\Omega =0.1$ eV), the relevant physical parameters are $m_{e,h} = 8.7\times 10^{-5}m_0, \upsilon_{Fx}= 0.208 \upsilon_F,  E_m =  \frac{-2\times 10^{-6}}{(m-1/2)^2}E_{H}$ and $\tilde{a}_{e}= \epsilon_b/\tilde{\mu} =1.14\times 10^5 $ for $\epsilon_b = 4.58$ for hBn. This shows that in the presence of  circularly polarized light, Kek$-$Y graphene supports weakly bound excitons in compared with any other gapped semiconductor because of its very light Dirac particles. \cite{ioffe}

\medskip
\par
With these dimensionless parameters, the expressions for the effective potential and two coupled differential equations turn out to be

\begin{equation}
\tilde{Z}_{m,p} (\tilde{X}) = 2 \tilde{m}_{e,h}  \int_{-\infty}^{\infty} d\tilde{x}\   \phi_{m}^{\ast}(\tilde{x})\phi_{p}(\tilde{x}) \left\{ \tilde{V}_{e}(\tilde{X}- \tilde{x}/2 ) + \tilde{V}_{h}(\tilde{X}+\tilde{x}/2 )  \right\}\  ,
\label{eff_dimnles}
\end{equation}

 \begin{eqnarray}
\frac{d R_{mn}(\tilde{y})}{d\tilde{y}} &= &-\sum_{j=0}^{\infty} \frac{1}{2i \tilde{K}_{j}^{\prime}} \left\{e^{i\left(\tilde{K}_{j}^{\prime}-\frac{3\tilde{m}_{e,h}\tilde{\upsilon}_{Fx}}{2}\right)\tilde{y}} \delta_{mj} + R_{mj}(\tilde{y})e^{-i\left(\tilde{K}_{j}^{\prime}-\frac{3\tilde{m}_{e,h}\tilde{\upsilon}_{Fx}}{2}\right)\tilde{y}} \right\} \nonumber\\
&\times& \sum_{p=0}^{\infty}\tilde{Z}_{jp}(\tilde{y}) \left\{e^{i\left(\tilde{K}_{p}^{\prime}-\frac{3\tilde{m}_{e,h}\tilde{\upsilon}_{Fx}}{2}\right)\tilde{y}} \delta_{pn} + R_{pn}(\tilde{y})e^{-i\left(\tilde{K}_{p}^{\prime}-\frac{3\tilde{m}_{e,h}\tilde{\upsilon}_{Fx}}{2}\right)\tilde{y}} \right\}, 
\label{diff_ref}
\end{eqnarray}
and 

\begin{eqnarray}
\frac{d T_{mn}(\tilde{y})}{d\tilde{y}} &= &-\sum_{j=0}^{\infty} \frac{1}{2i \tilde{K}_{j}^{\prime}} \left\{e^{i\left(\tilde{K}_{j}^{\prime}-\frac{3\tilde{m}_{e,h}\tilde{\upsilon}_{Fx}}{2}\right)\tilde{y}}   T_{mj}(\tilde{y}) \right\} \nonumber\\
&\times& \sum_{p=0}^{\infty}\tilde{Z}_{jp}(\tilde{y}) \left\{e^{i\left(\tilde{K}_{p}^{\prime}-\frac{3\tilde{m}_{e,h}\tilde{\upsilon}_{Fx}}{2}\right)\tilde{y}} \delta_{pn} + R_{pn}(\tilde{y})e^{-i\left(\tilde{K}_{p}^{\prime}-\frac{3\tilde{m}_{e,h}\tilde{\upsilon}_{Fx}}{2}\right)\tilde{y}} \right\} 
\label{diff_trans}
\end{eqnarray}
where, 
\begin{equation}
\phi_p(\tilde{x}) = N_p \frac{2\tilde{x}}{(p+1) \tilde{a}_e} e^{-|\frac{2\tilde{x}}{(p+1) \tilde{a}_e}|/2} L_{p+1}^1 (|\frac{2\tilde{x}}{(p+1) \tilde{a}_e}|) \  .
\label{eign_fun_exciton}
\end{equation}

\medskip
\par

 \begin{figure}
\centering
\includegraphics[width=0.45\textwidth]{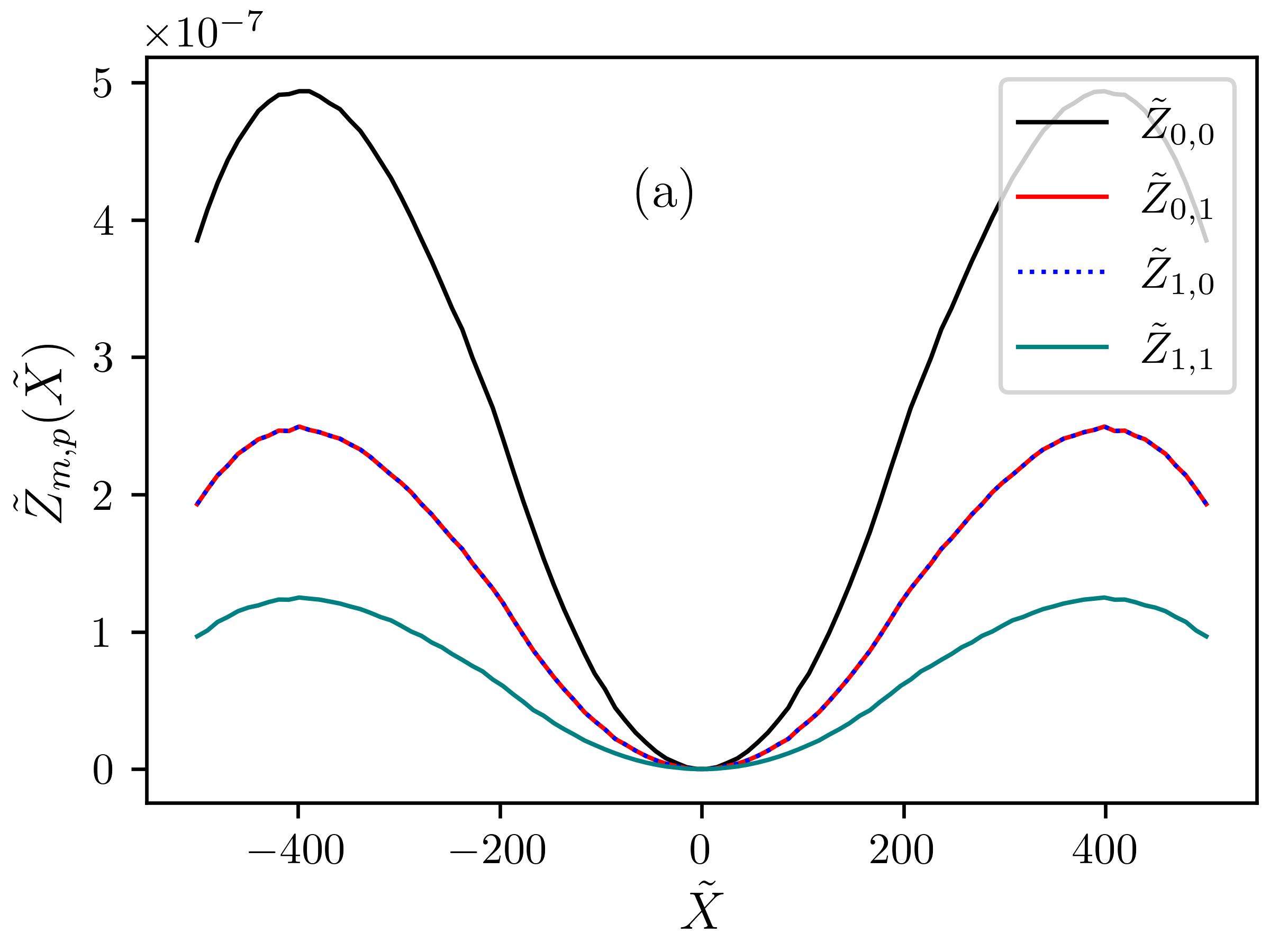}
\includegraphics[width=0.45\textwidth]{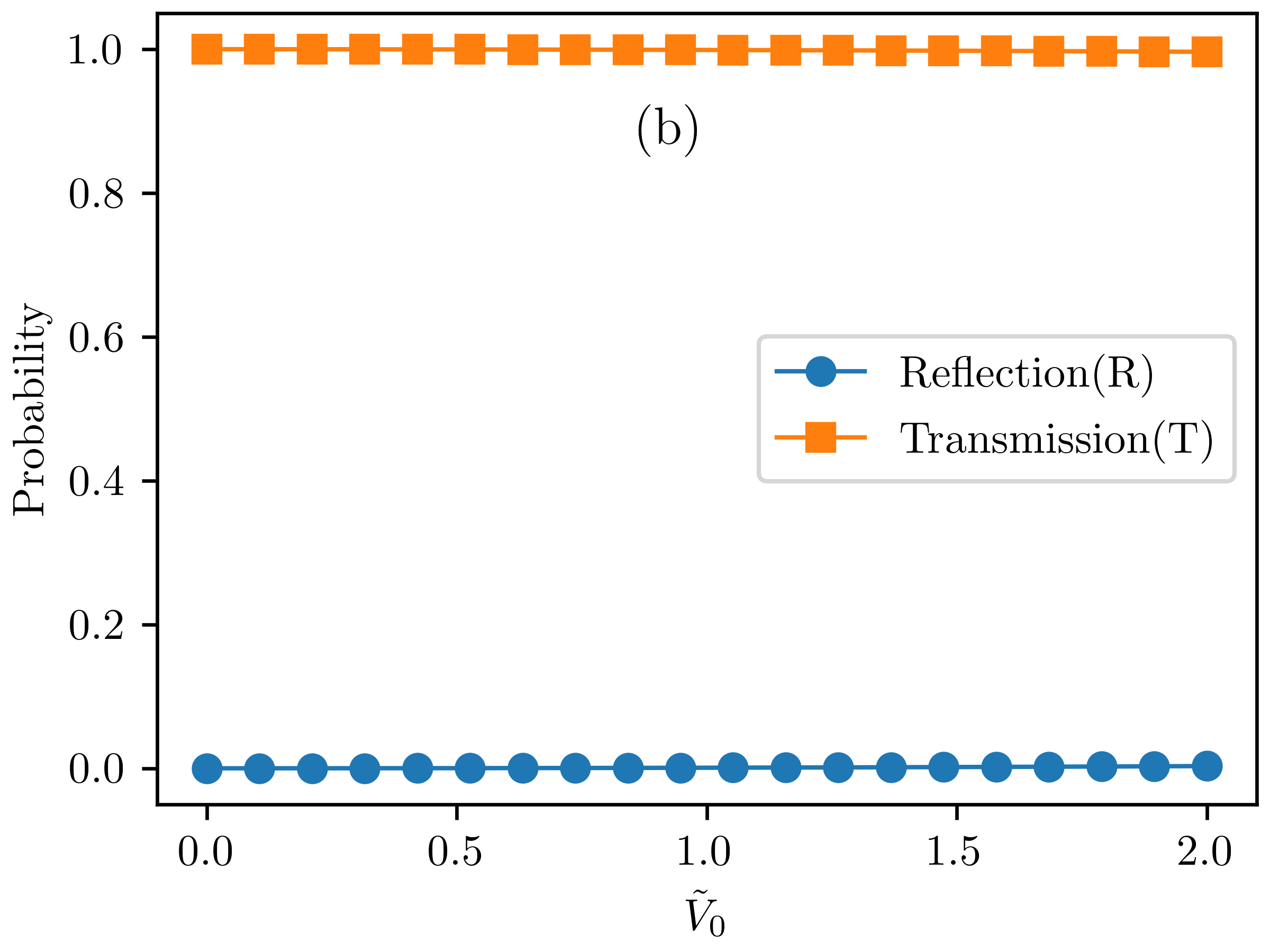}
\caption{(Color online) (a) Plot of the effective potential $\tilde{Z}_{m,p}(\tilde{X})$ as a function of the center of mass coordinate $\tilde{X}$ for the center of mass  of  the exciton when incident in the state $p$ and reflected or transmitted in the state $m$ across a square potential barrier of width $d$ and strength $V_0$. (b)  Plot of the probability of transmission or reflection of an exciton in state $n=0$ across a square potential barrier as a function of the strength of the barrier. In the plots, all parameters are dimensionless. We have used $\tilde{m}_{e,h} =8.7\times 10^{-5} $  which is for irradiation parameter $\tilde{\zeta} = 0.1, \hbar\Omega = 0.1$ eV and $\upsilon_F = 10^{6}$m/s. Other parameters include $\epsilon_b = 4.58, \tilde{a}_e= 2\epsilon_b/\tilde{m_{e,h}}$ and , $\tilde{d} =100 $. It is noted that the strengths of the external potential barrier for electron and hole are equal and we have used  twice that of the binding energy of  the hydrogen atom in (a).  In plot (b), the kinetic energy of  the exciton is $2\times10^{-4}E_{H} \to 2.7$ meV and the probability we have plotted is for potential $0$ to $10^4 E_y$. }
\label{FIG:9}
\end{figure}

\medskip
\par
\begin{figure}
\centering
\includegraphics[width=0.45\textwidth]{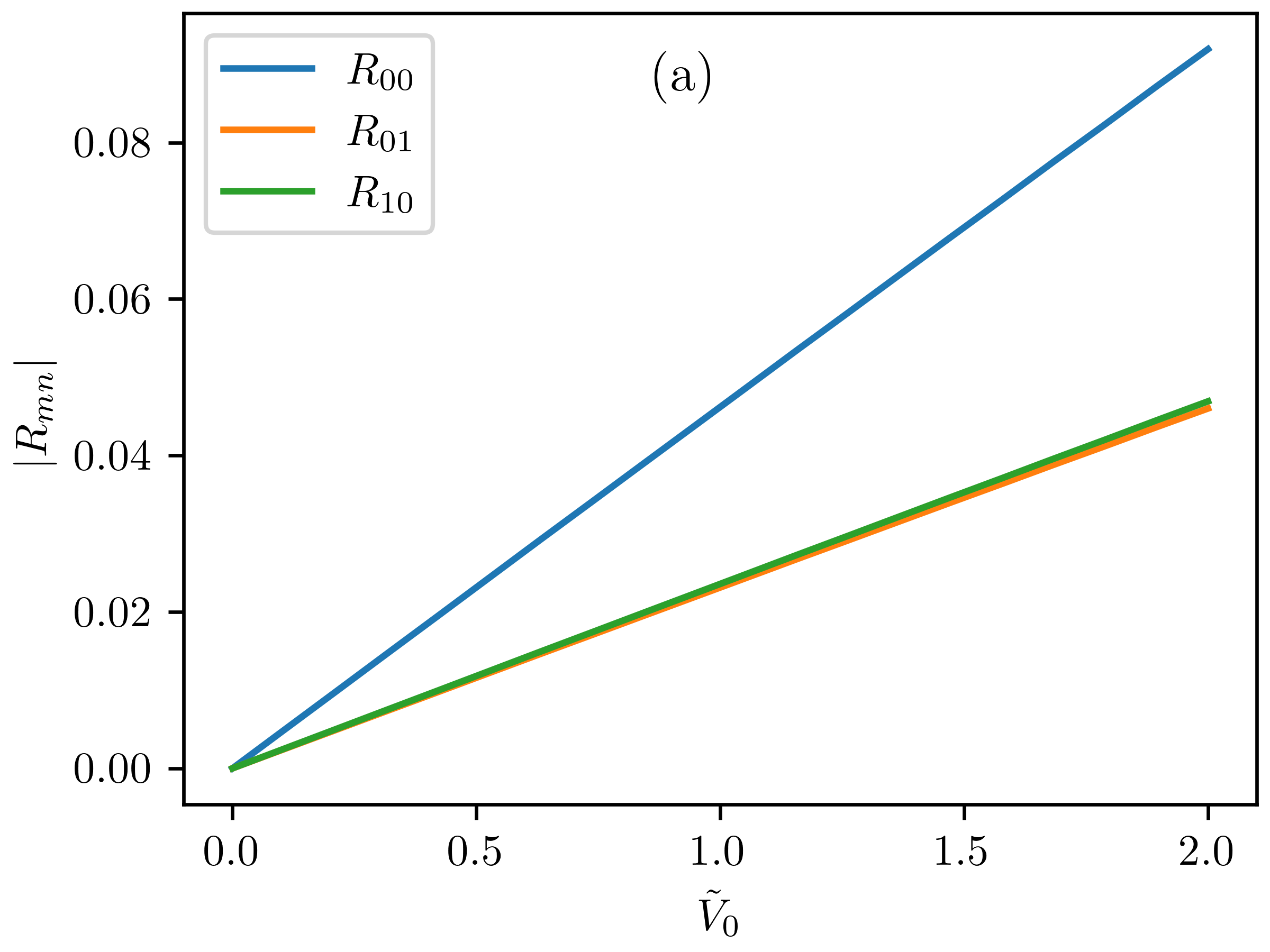}
\includegraphics[width=0.45\textwidth]{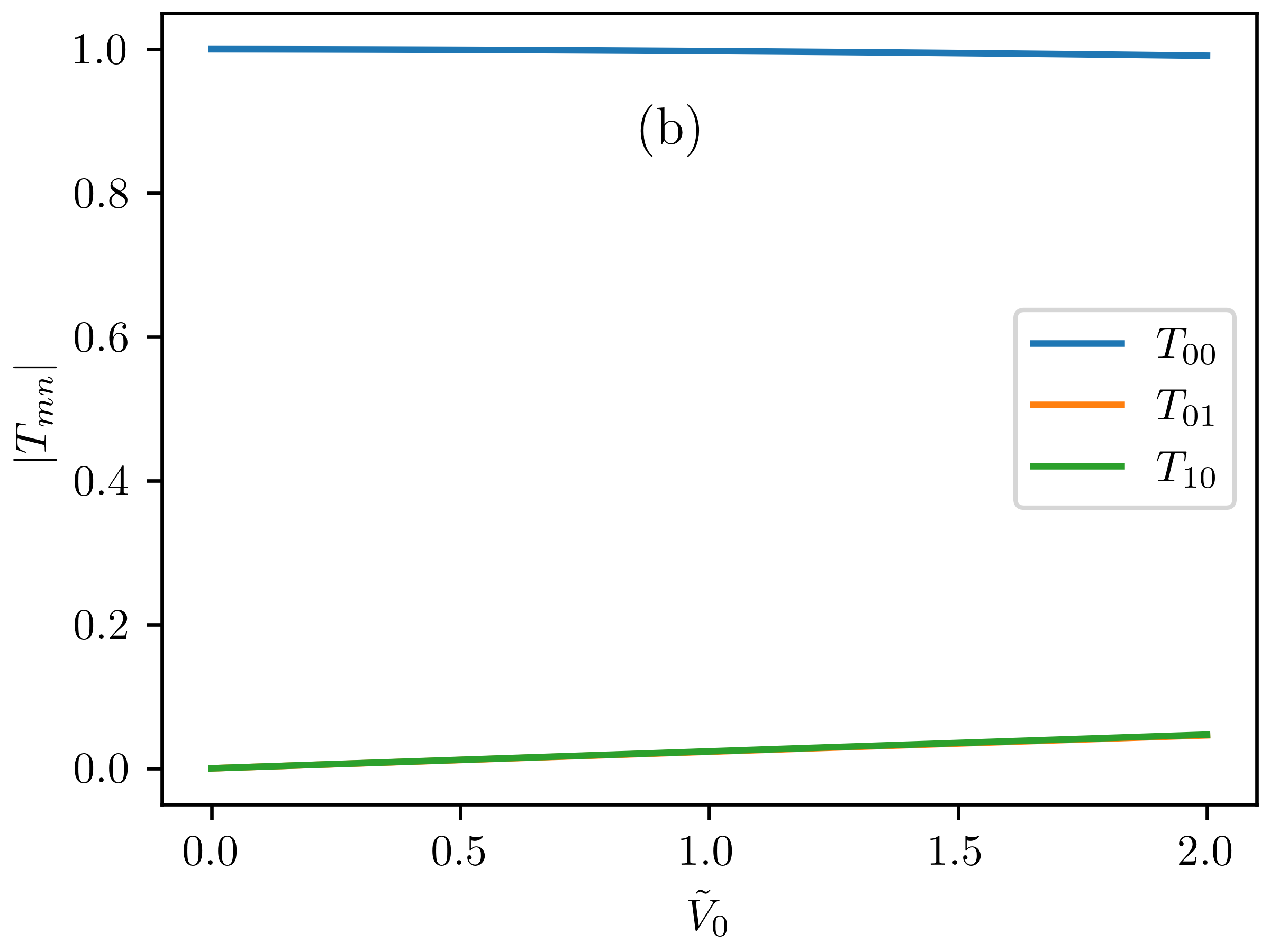}
\caption{(Color online) Plot of the probability amplitude for the center of mass  of  the exciton  as a function of the potential height $\tilde{V}_0$ of a finite width square potential barrier. The exciton is incident in state $n$ and reflected or transmitted in state $m$. Plot  (a) is for the reflection amplitude and plot (b) is for the transmission amplitude. In the plot all parameters are dimensionless. We chose $\tilde{m}_{e,h} =8.7\times 10^{-5}  $  which is for irradiation parameter $\tilde{\zeta} = 0.1, \hbar\Omega = 0.1$ eV and $\upsilon_F = 10^{6}$m/s. Other parameters include, $\epsilon_b = 4.58, \tilde{a}_e= 2\epsilon_b/\tilde{m_{e,h}}$ and , $\tilde{d} =100 $. It is noted that the strengths of the external potential barrier for the electron and hole are equal and we chose the exciton kinetic energy as $2\times10^{-4}E_{H} \to 2.7$ meV and we have plotted the probability for potential $0$ to $10^4 E_y$. }
\label{FIG:10}
\end{figure}

\medskip
\par

First, we consider a square potential barrier  $V_0\,  \,  \, \text{for}\ 0\leq x_{e,h}\leq  d$ and zero otherwise.    We plot the center of mass effective potential as a function of center of mass coordinate for equal strength $V_0 = 2\times E_{H}$ for electron and hole in Fig.\, \ref{FIG:9}(a). It is observed that the effective potential  experienced by the center of mass of the exciton is negligible or it almost vanishes and hence perfect transmission is obtained in Fig.\, \ref{FIG:9}(b) for finite range of the potential strength. This numerical result can be proved analytically from the expression for effective potential $\tilde{Z}_{m,p}(\tilde{X})$ in Eq.\, (\ref{eff_dimnles}), where the effective potential depends not only on the strength of the external barrier but also on the effective mass and the probability that the exciton incident state on the left-hand side of the barrier and on the outgoing state on the right-hand side of the barrier. Since the exciton effective mass is very small $\sim 10^{-5} m_{0}$ in terms of the free electron rest mass  it suppress the integrand transition probability over  the potential strength. Therefore, the effective potential comes out to be of order $10^{-7}E_{H}$  until the potential strength is infinitely large with $V_0 > 10^6$ eV to overcome the role of the effective mass. Hence perfect transmission is observed for the large but finite range of potential strength. It should be mentioned that these numerical results are for barrier width $d=100\, a_{0}$ which is smaller than the enormously large Bohr radius of an exciton. But for very small effective mass exciton even when the barrier is wider than the exciton Bohr radius the average scalar potential seen by the center of mass is infinitesimally small. Additionally, the probability of transmission of an exciton from one state to another is also destroyed and hence perfect transmission is only allowed between the same eigenstate as if there is no obstacle  in the path of propagation of the center of mass of an exciton. This conclusion is based on the numerical result presented in Fig.\, \ref{FIG:10}, where the amplitude of reflection and transmission between  the ground state and the first excited state are plotted as a function of the potential strength. Analytically, it is because the limits of integration for $x$ are symmetric.  However, because of orthogonality of the wave functions $\phi_{m,p}(x)$, the integrand is even only when $m=p$. Therefore, the off-diagonal elements are automatically cancelled. From this discussion, it is clear that the effective center of mass potential of a light but enormous Dirac exciton vanishes identically in a scalar barrier of finite height. And hence these excitons incident on a square potential barrier is transmitted to the other side of the barrier without any obstruction.  

\medskip
\par

\begin{figure}
\centering
\includegraphics[width=0.45\textwidth]{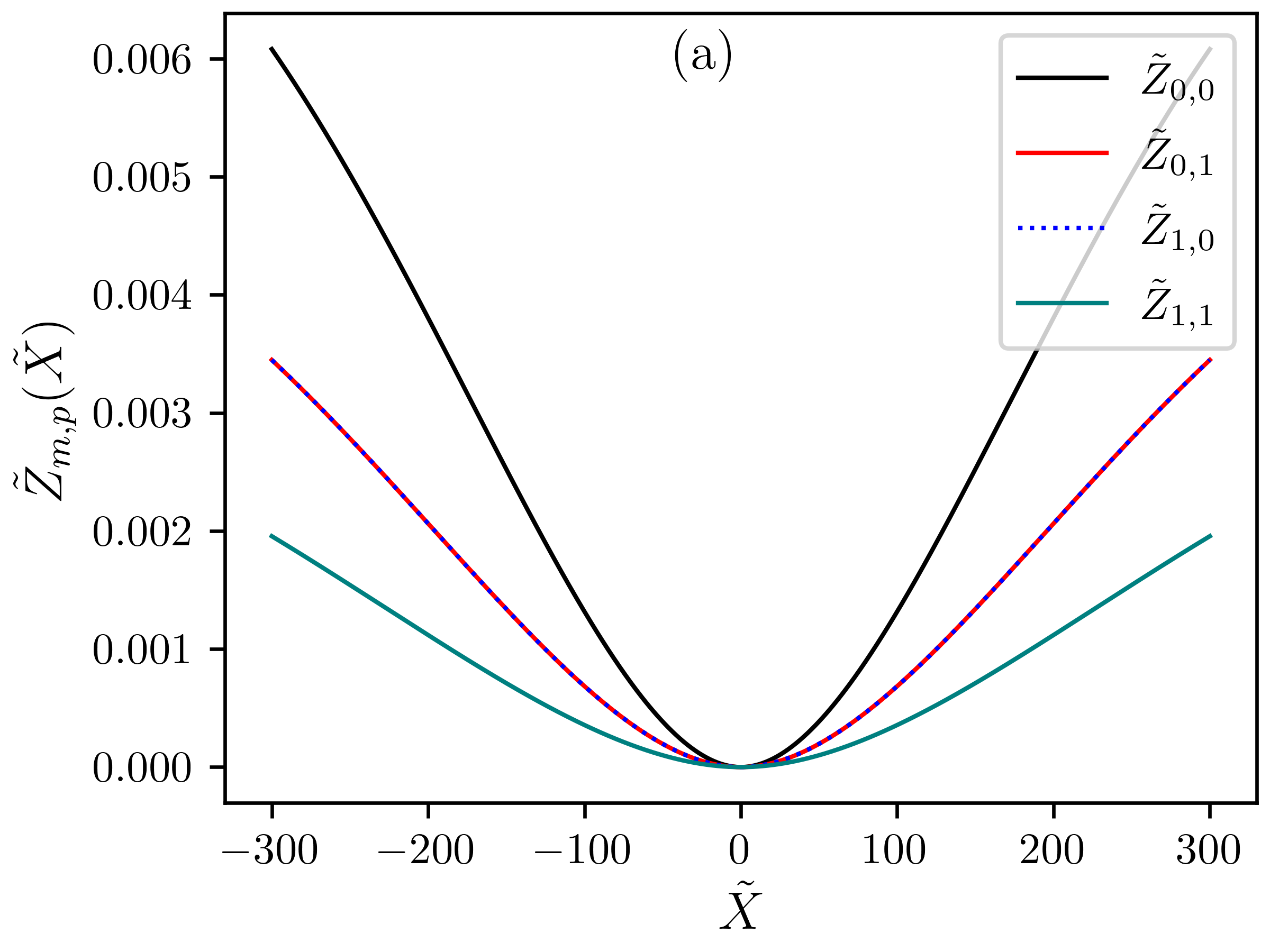}
\includegraphics[width=0.45\textwidth]{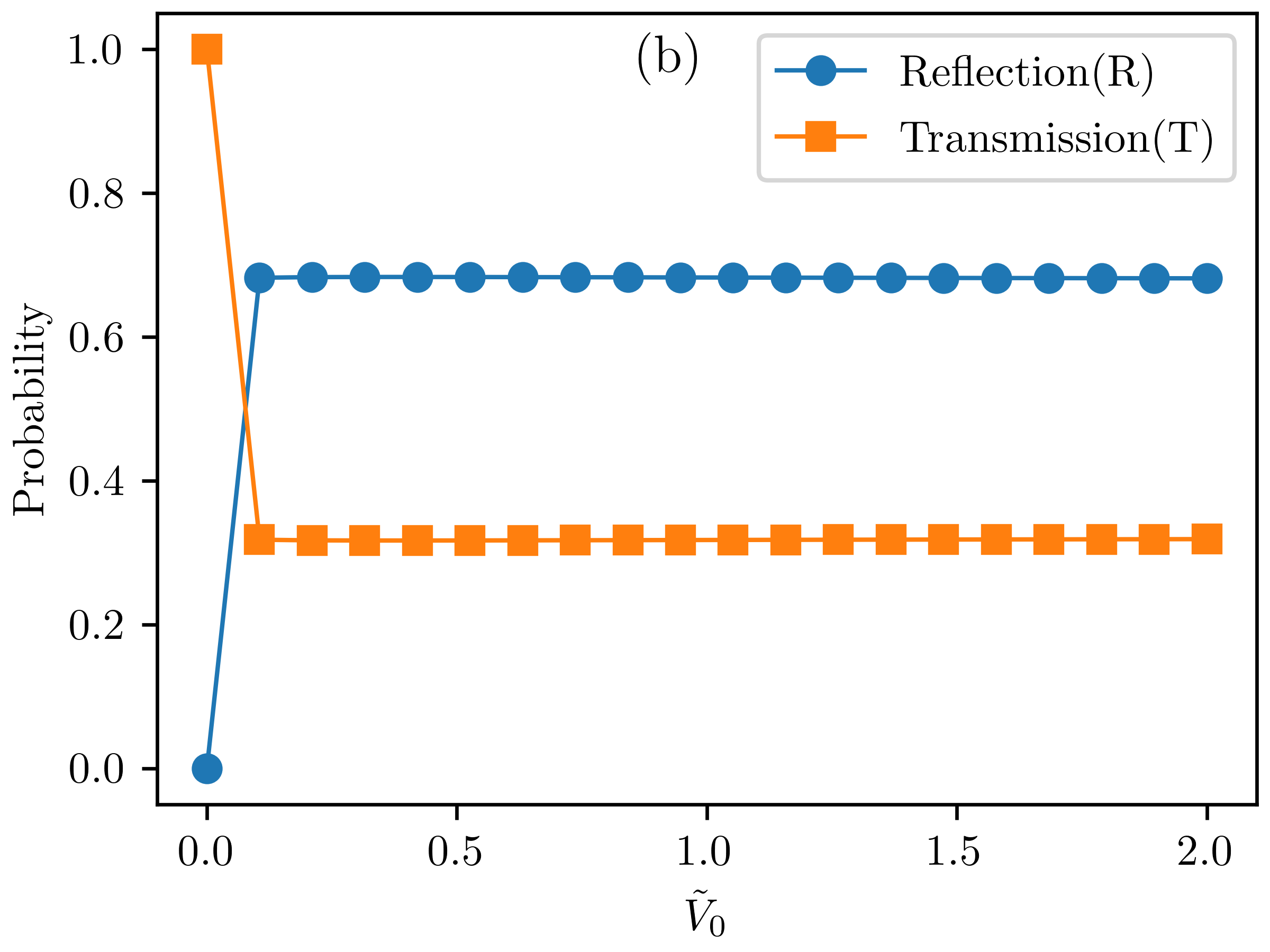}
\caption{(Color online) (a)  Plot of the effective potential $\tilde{Z}_{m,p}(\tilde{X})$ as a function of the center of mass coordinate $\tilde{X}$ for the center of mass  of  the exciton when incident in state $p$ and reflected or transmitted in the state $m$ through a delta potential barrier or unequal strength $V_{0;e}$ and $V_{0;h}$, where $V_{0;e} = V_0 $ and $V_{0;h} = 10 V_0 $ . (b)  Plots of the probability of transmission and reflection of an exciton in state $n=0$ across a square potential barrier as a function of height of the barrier. In the plots, all parameters are dimensionless. We have chosen $\tilde{m}_{e,h} =3.5\times 10^{-3}  $ which is for irradiation parameter $\tilde{\zeta} = 0.2, \hbar\Omega = 1$ eV and $\upsilon_F = 10^{6}$m/s. Other parameters include, $\epsilon_b = 4.58, \tilde{a}= 2\epsilon_b/\tilde{m_{e,h}}$ and  $\tilde{d} =100 $. It is noted that the strength of the delta potential barrier is  twice that of the binding energy of  the  hydrogen atom in (a).  In plots  (b), the kinetic energy of an exciton is $2\times10^{-4}E_{H} \to 2.7$meV and the probability we have plotted for $V_0$ is $0$ to $10^4 E_y$. }
\label{FIG:11}
\end{figure}

 While discussing these results we have to keep in mind that, the electron-hole quasiparticles are  spatially separated   in the presence of the externally applied circularly polarized high frequency irradiation. Additionally, the irradiation induced effective mass of electron and holes are equal and can be tuned by manipulating the irradiation parameter but we cannot choose the parameters arbitrarily just to obtain a heavy exciton. We have also assumed that the heights of the external potential barrier for the electron and hole are equal. If one considers the irradiation parameter $\tilde{\zeta}$ high but still less than $1$, and the frequency is increased by a factor of  $10$,  the effective mass can be increased by some amount. For the larger effective mass and unequal potential for the electron and hole, the exciton may have finite reflection probability.  In Fig.\, \ref{FIG:11}, numerical results are presented for effective potential and probabilities when the effective mass of the electron and hole $m_{e,h} =3.5\times 10^{-3} m_{0} $,  group velocity $\upsilon_{Fx} = 0.416\upsilon_0$ and delta potential of unequal strength for electron and hole given by $V_{e,h}(x) = V_{0;e,h}\delta(x)$. Here,  the effective mass and the group velocity are calculated for irradiation parameter $\tilde{\zeta}=0.2$ and for photon of energy 1 eV. 
\medskip
\par

\medskip
\par

The results in Fig.\,  \ref{FIG:11}(a), show that the center of mass of the exciton experienced a very small potential than the individual electron and hole.   Additionally, the effective potential depends not only on the strength of the external potential but also the probability for the incident state of the exciton on the left-hand side of the potential and the outgoing state on the right-hand side  of the potential. For example,  the actual potential seen by the exciton incident in the ground state and transmitted or reflected to the ground state $\tilde{Z}_{0,0}$ is higher than the absolute potential seen by the ground state exciton transmitted or reflected to the first excited state $\tilde{Z}_{1,0}$. Therefore, the exciton in the ground state has higher probability for reflection to the ground state than reflected to the first excited state.  It is also seen that the exciton in the ground state and transmitted or reflected to the first excited state have equal effective potential for the exciton in the first excited state transmitted or reflected to the ground state. i.e $\tilde{Z}_{1,0}=\tilde{Z}_{0,1}$.  However the total probability of the quasiparticle from ground state cross the potential barrier is the sum of the probability of transmission from ground state to any possible state. The  Fig. \ref{FIG:11}(b) explain that perfect transmission is no longer present when the potential strength of the electron and holes are higher than that of the kinetic energy of the incident exciton. Even the effective potential the center of mass encounters is small it is still greater than the incident kinetic energy of exciton. Therefore quasiparticles are partially transmitted and the transmission probability is constant of potential strength.

 \section{Concluding Remarks}
\label{sec6}
We have investigated the effect due to high frequency irradiation on valley resolved transmission and reflection of charged particles incident on a square potential barrier on monolayer  Kek$-$Y distorted graphene. We employed a low energy Hamiltonian  near the Dirac points of Kekul\'e distorted graphene lattice with Y-shaped bond texture. To this, we added a high frequency time-dependent perturbation due to external irradiation. We employed the Floquet Magnus expansion to obtain a time independent effective Hamiltonian which has the same stroboscopic effect as the  applied time dependent perturbative Hamiltonian for elliptically polarized irradiation. We determined the dressed states and modified energy spectra due to irradiation. The results show that, a dynamical gap between the conduction and valance bands is induced as well as two folded cones are separated by small energy gap at the Dirac point due to circularly polarized irradiation.

 \medskip
\par

 After establishing our model Hamiltonian,  our next step was to calculate the probability current of transmission and reflection for both undressed and dressed states by solving the continuity equations which apply to the wave functions.  Maintaining the translational symmetry along the transverse direction and employing the boundary conditions at the edge of the barrier, an analytical expression for the valley dependent transmission and reflection coefficients were obtained. The presented analytical and  numerical results for the transmission probability separating the intervalley and intravalley contributions  demonstrates that valley manipulation for the transmission can enhance the Klein tunneling at around normal incidence at small barrier width and suppress resonant tunneling when  the barrier becomes wider and wider. Additionally, the electrical conductance is decreased due to opening  the new channel (intervalley) for scattering. This suggests that valley degree of freedom can be used in a similar way as spin to control electron transmission through the potential region.   Furthermore, highly suppressed transmission probability for irradiated electrons at high frequency, supports the idea that illuminating the material with high frequency electromagnetic field is another effective method for  modifying the electronic spectra and controlling the transport properties of graphene. The polar plots for the transmission probability for chosen ratio of irradiation intensity to the frequency, demonstrate that the effect of irradiation is comparatively more  significant on intervalley scattering than on the intravalley scattering.

\medskip
\par

Formation of excitons due to a dynamical gap between electron and hole subbands is another consequence  of irradiation which established the ground for the study of the transport properties of such bound quasiparticles in Kek$-$Y graphene.  First, we calculated the electronic spectra of  excitons following the one dimensional hydrogenic model and  obtained an expression for the effective mass  and group velocity of electron and hole in both valleys.  The effective mass of an electron and hole in the outer cone is found to be divergent while that for inner cone is equal. As a consequence the intervalley transmission of excitons are not possible in our formalism. Therefore we  calculated  the reflection and transmission probabilities of an exciton from inner cone incident upon a square potential barrier of equal strength for electron and hole using the numerical method of variable transmission and reflection amplitudes. Our results demonstrate that there is perfect tunneling of an exciton from the ground state of the inner cone through the symmetric potential barrier of finite height irrespective of the height of the barrier as the effective  potential experienced by the center of mass of the  exciton is insignificantly small in comparison to the potential experienced by the individual electron and hole.  Moreover, the even and odd parity coupling between exciton states is forbidden due to the conservation of parity and the symmetry in the Hamiltonian for electron and hole. When a symmetric potential barrier is replaced by a delta potential barrier of unequal strength, the perfect transmission is broken and hence the significant reflection is observed for finite strength of the  potential.

\medskip
\par
In conclusion, high frequency circularly polarized irradiation  significantly modifies the electronic spectra and offers more appealing transport phenomena in Kek$-$Y distorted graphene including suppressed transmission and perfect tunneling of excitons. Another important implication is that these properties can be tuned by varying the values for the intensity and frequency of the irradiation. Our analytical and numerical results correctly reproduce that of unirradiated Kek$-$Y graphene in the limit of irradiation parameter $\zeta $ tends to zero. We believe that our theoretical framework and the numerical analysis are experimentally feasible  and these results have their own application in designing and manufacturing the electronic, optoelectronic and valleytronic devices based on graphene.

\medskip
\par
\begin{acknowledgements}
G.G. gratefully acknowledges funding from the U.S. National Aeronautics and Space Administration (NASA) via the NASA-Hunter College Center for Advanced Energy Storage for Space  under cooperative agreement 80NSSC24M0177.  We also acknowledge support from the Science and Technology Facilities Council (STFC), UK (Reference No. ST/Y005147/1)
\end{acknowledgements}

\medskip
\par

\appendix

\section{Numerical solution of the Transmission  coefficient}

A lengthy calculation is needed for solving the equations (\ref{eq_boundry}) analytically.    So, we turn to solving these equations numerically by constructing them in matrix form as, $ML=W$, where $M$ is given by 

\begin{equation}
 M
   = \  \left(
 \begin{array}{cccccccc}
 -e^{i\theta_{k^{+}}} & -e^{i\theta_{k^{-}}} &-e^{-i\theta_{q^{+}}}&-e^{-i\theta_{q^{-}}} &e^{i\theta_{q^{+}}} &e^{i\theta_{q^{-}}} &0&0\\
 s&s&-s^{\prime}&-s^{\prime}&-s^{\prime}&-s^{\prime}&0&0\\
 s&-s&-s^{\prime}&s^{\prime}&-s^{\prime}&s^{\prime}&0&0\\
 -e^{-i\theta_{k^{+}}} & e^{-i\theta_{k^{-}}} &-e^{i\theta_{q^{+}}}&e^{i\theta_{q^{-}}} &e^{-i\theta_{q^{+}}} &-e^{-i\theta_{q^{-}}} &0&0\\
 0&0&e^{i(q^{+}_xd -\theta_{q^{+}})} &e^{i(q^{-}_xd -\theta_{q^{-}})} &-e^{-i(q^{+}_xd -\theta_{q^{+}})} &-e^{-i(q^{-}_xd -\theta_{q^{-}})} &-e^{i(k^{+}_xd -\theta_{k^{+}})} &-e^{i(k^{-}_xd -\theta_{k^{-}})} \\
 0&0&s^{\prime}e^{iq^{+}_xd}&s^{\prime}e^{iq^{-}_xd}&s^{\prime}e^{-iq^{+}_xd}&s^{\prime}e^{-iq^{-}_xd}&-s e^{ik^{+}_xd}&-s e^{ik^{-}_xd}\\
 0&0&s^{\prime}e^{iq^{+}_xd}&-s^{\prime}e^{iq^{-}_xd}&s^{\prime}e^{-iq^{+}_xd}&-s^{\prime}e^{-iq^{-}_xd}&-s e^{ik^{+}_xd}&s e^{ik^{-}_xd}\\
 0&0&e^{i(q^{+}_xd +\theta_{q^{+}})} &-e^{i(q^{-}_xd +\theta_{q^{-}})} &-e^{-i(q^{+}_xd +\theta_{q^{+}})} &-e^{-i(q^{-}_xd +\theta_{q^{-}})} &-e^{i(k^{+}_xd +\theta_{k^{+}})} &e^{i(k^{-}_xd +\theta_{k^{-}})} 
 \end{array}
 \right) 
 \label{m_matrix}
\end{equation}
and $L$ and $W$ are column vectors with transpose:  

\begin{equation}
L=  \left(
\begin{array}{cccccccc}
r_+&r_-&\gamma_+&\gamma_-&\eta_+&\eta_-&t_+&t_- 

\end{array}
\right)^T 
\end{equation}

\begin{equation}
W=  \left(
\begin{array}{cccccccc}
-e^{-i\theta_{k^{+}}} & -s & -s & -e^{i\theta_{k^{+}}} & 0 & 0 & 0 & 0 
\end{array}
\right)^T \  .
\end{equation} 
Using Cramer's rule for the $j^{th}$ unknown, $x_j =$det$(M_j)/$det$(M)$ where $M_j$ is the matrix obtained from $M$ by replacing its $j^{th}$ column by $W$,   We obtained $r_+ =$det$(M_1)/$det$(M)$, $r_-=$det$(M_2)/$det$(M)$ , $t_+ =$det$(M_7)/$det$(M)$ and $t_- =$det$(M_8)/$det$(M)$.

\medskip
\par

\section{Transmission subjected to irradiation}

The reflected and transmitted current from each valley  can be expressed as, 

\begin{eqnarray}
J_x^{({\tilde r}_+)} &=& -2\upsilon_F \frac{|{\tilde r}_+|^2}{(1+ \text{b}^2_s(\pi-\theta_{k^{+}}))(1+ \text{b}^2_+(\pi-\theta_{k^+}))} \left[ \cos\theta_{k^{+}} \{\text{b}_s(\pi-\theta_{k^{+}})(1+\text{b}^2_+(\pi-\theta_{k^{+}}))  +\Delta_0 \text{b}_+(\pi-\theta_{k^{+}})  (1+\text{b}^2_s(\pi-\theta_{k^{+}}) \} \right.\nonumber\\
& &\left. -\tilde\zeta \{  (1-\text{b}^2_s(\pi-\theta_{k^{+}}))(1+\text{b}^2_+(\pi-\theta_{k^{+}}))+\Delta_0^2 (1+\text{b}^2_s(\pi-\theta_{k^{+}}))(1-\text{b}^2_+(\pi-\theta_{k^{+}}))  \}   \right],
\label{jx_rp_tild}
\end{eqnarray}

\begin{eqnarray}
J_x^{({\tilde r}_-)} &=& -2\upsilon_F \frac{|{\tilde r}_-|^2}{(1+ \text{b}^2_s(\pi-\theta_{k^{-}}))(1+ \text{b}^2_-(\pi-\theta_{k^-}))} \left[ \cos\theta_{k^{-}} \{\text{b}_s(\pi-\theta_{k^{-}})(1+\text{b}^2_-(\pi-\theta_{k^{-}}))  +\Delta_0 \text{b}_-(\pi-\theta_{k^{-}})  (1+\text{b}^2_s(\pi-\theta_{k^{-}}) \} \right.\nonumber\\
& &\left. -\tilde\zeta \{  (1-\text{b}^2_s(\pi-\theta_{k^{-}}))(1+\text{b}^2_-(\pi-\theta_{k^{-}}))+\Delta_0^2 (1+\text{b}^2_s(\pi-\theta_{k^{-}}))(1-\text{b}^2_-(\pi-\theta_{k^{-}}))  \}   \right],
\label{jx_rm_tild}
\end{eqnarray}

\begin{eqnarray}
J_x^{({\tilde t}_+)} &=& 2\upsilon_F \frac{|{\tilde t}_+|^2}{(1+ \text{b}^2_s(\theta_{k^{+}}))(1+ \text{b}^2_+(\theta_{k^+}))} \left[ \cos\theta_{k^{+}} \{\text{b}_s(\theta_{k^{+}})(1+\text{b}^2_+(\theta_{k^{+}}))  +\Delta_0 \text{b}_+(\theta_{k^{+}})  (1+\text{b}^2_s(\theta_{k^{+}}) \} \right.\nonumber\\
& &\left. +\tilde\zeta \{  (1-\text{b}^2_s(\theta_{k^{+}}))(1+\text{b}^2_+(\theta_{k^{+}}))+\Delta_0^2 (1+\text{b}^2_s(\theta_{k^{+}}))(1-\text{b}^2_+(\theta_{k^{+}}))  \}   \right],
\label{jx_tp_tild}
\end{eqnarray}
and 

\begin{eqnarray}
J_x^{({\tilde t}_-)} &=& 2\upsilon_F \frac{|{\tilde t}_-|^2}{(1+ \text{b}^2_s(\theta_{k^{-}}))(1+ \text{b}^2_-(\theta_{k^-}))} \left[ \cos\theta_{k^{-}} \{\text{b}_s(\theta_{k^{-}})(1+\text{b}^2_-(\theta_{k^{-}}))  +\Delta_0 \text{b}_+(\theta_{k^{-}})  (1+\text{b}^2_s(\theta_{k^{-}}) \} \right.\nonumber\\
& &\left. +\tilde\zeta \{  (1-\text{b}^2_s(\theta_{k^{-}}))(1+\text{b}^2_-(\theta_{k^{-}}))+\Delta_0^2 (1+\text{b}^2_s(\theta_{k^{-}}))(1-\text{b}^2_-(\theta_{k^{-}}))  \}   \right],
\label{jx_tm_tild}
\end{eqnarray}
with the corresponding reflection and transmission wave  functions   as follows:

\begin{equation}
\label{wave_rp_c}
\Psi^{E({\tilde r}_+)}_{Y}( {\bf k};{\bf r}) = \  \frac{{\tilde r}_+}{\sqrt{(1+ \text{b}^{2}_{s}(\pi- \theta_{k^{+}}))(1+\text{b}^{2}_{+}(\pi-\theta_{k^{+}}) )}} \, \left(
\begin{array}{c}
-\text{e}^{ i \theta_{k^{+}}} \\
\text{b}_s(\pi- \theta_{k^{+}})  \\
\text{b}_+(\pi- \theta_{k^{+}})  \\
-\text{b}_s(\pi-\theta_{k^{+}})  \text{b}_+(\pi- \theta_{k^{+}}) \text{e}^{- i \theta_{k^+}}          
\end{array}
\right)
\frac{e^{ik_y y - i| k^+_x |x}}{\sqrt{A}}\  ,   
\end{equation}

\begin{equation}
\label{wave_rm_c}
\Psi^{E({\tilde r}_-)}_{Y}( {\bf k};{\bf r}) = \  \frac{{\tilde r}_-}{\sqrt{(1+ \text{b}^{2}_{s}(\pi- \theta_{k^{-}}))(1+\text{b}^{2}_{-}(\pi-\theta_{k^{-}}) )}} \, \left(
\begin{array}{c}
-\text{e}^{ i \theta_{k^{-}}} \\
\text{b}_s(\pi- \theta_{k^{-}})  \\
\text{b}_-(\pi- \theta_{k^{-}})  \\
-\text{b}_s(\pi-\theta_{k^{-}})  \text{b}_-(\pi- \theta_{k^{-}}) \text{e}^{- i \theta_{k^-}}          
\end{array}
\right)
\frac{e^{ik_y y - i| k^-_x |x}}{\sqrt{A}}\  ,    
\end{equation}

\begin{equation}
\label{wave_tp_c}
\Psi^{E({\tilde t}_+)}_{Y}( {\bf k};{\bf r}) = \  \frac{{\tilde t}_+}{\sqrt{(1+ \text{b}^{2}_{s}(\theta_{k^{+}}))(1+\text{b}^{2}_{+}(\theta_{k^{+}}) )}} \, \left(
\begin{array}{c}
\text{e}^{- i \theta_{k^{+}}} \\
\text{b}_s( \theta_{k^{+}})  \\
\text{b}_+(\theta_{k^{+}})  \\
\text{b}_s(\theta_{k^{+}})  \text{b}_+( \theta_{k^{+}}) \text{e}^{ i \theta_{k^+}}          
\end{array}
\right)
\frac{e^{ik_y y +i| k^+_x |x}}{\sqrt{A}} \  ,    
\end{equation}
 and 

\begin{equation}
\label{wave_tm_c}
\Psi^{E({\tilde t}_-)}_{Y}( {\bf k};{\bf r}) = \  \frac{{\tilde t}_-}{\sqrt{(1+ \text{b}^{2}_{s}(\theta_{k^{-}}))(1+\text{b}^{2}_{-}(\theta_{k^{-}}) )}} \, \left(
\begin{array}{c}
\text{e}^{- i \theta_{k^{-}}} \\
\text{b}_s( \theta_{k^{-}})  \\
\text{b}_+(\theta_{k^{-}})  \\
\text{b}_s(\theta_{k^{-}})  \text{b}_-( \theta_{k^{-}}) \text{e}^{ i \theta_{k^-}}          
\end{array}
\right)
\frac{e^{ik_y y +i| k^-_x |x}}{\sqrt{A}}\  .    
\end{equation}

\medskip
\par
In the scattering region  (II), the forward moving and backward moving waves are defined as follows,

\begin{equation}
\label{wave_fp_c}
\Psi^{E({\tilde P}_+)}_{Y}( {\bf k};{\bf r}) = \  \frac{{\tilde P}_+}{\sqrt{(1+ \text{b}^{2}_{s^{\prime}}( \theta_{q^{+}}))(1+\text{b}^{2}_{+}(\theta_{q^{+}}) )}} \, \left(
\begin{array}{c}
\text{e}^{-i \theta_{q^{+}}} \\
\text{b}_{s^{\prime}}( \theta_{q^{+}})  \\
\text{b}_+(\theta_{q^{+}})  \\
\text{b}_{s^{\prime}}(\theta_{q^{+}})  \text{b}_+( \theta_{q^{+}}) \text{e}^{i \theta_{q^+}}          
\end{array}
\right)
\frac{e^{ik_y y + i| q^+_x |x}}{\sqrt{A}}\  ,     
\end{equation}

\begin{equation}
\label{wave_fm_c}
\Psi^{E({\tilde P}_-)}_{Y}( {\bf k};{\bf r}) = \  \frac{{\tilde P}_-}{\sqrt{(1+ \text{b}^{2}_{s^{\prime}}( \theta_{q^{-}}))(1+\text{b}^{2}_{-}(\theta_{q^{-}}) )}} \, \left(
\begin{array}{c}
\text{e}^{-i \theta_{q^{-}}} \\
\text{b}_{s^{\prime}}( \theta_{q^{-}})  \\
\text{b}_-(\theta_{q^{-}})  \\
\text{b}_{s^{\prime}}(\theta_{q^{-}})  \text{b}_-( \theta_{q^{-}}) \text{e}^{i \theta_{q^-}}          
\end{array}
\right)
\frac{e^{ik_y y + i| q^-_x |x}}{\sqrt{A}}\  ,    
\end{equation}

\begin{equation}
\label{wave_bp_c}
\Psi^{E({\tilde Q}_+)}_{Y}( {\bf k};{\bf r}) = \  \frac{{\tilde Q}_+}{\sqrt{(1+ \text{b}^{2}_{s^{\prime}}(\pi-\theta_{q^{+}}))(1+\text{b}^{2}_{+}(\pi-\theta_{q^{+}}) )}} \, \left(
\begin{array}{c}
-\text{e}^{ i \theta_{q^{+}}} \\
\text{b}_{s^{\prime}}( \pi- \theta_{q^{+}})  \\
\text{b}_+(\pi- \theta_{q^{+}})  \\
-\text{b}_{s^{\prime}}(\pi- \theta_{q^{+}})  \text{b}_+(\pi- \theta_{q^{+}}) \text{e}^{ -i \theta_{q^+}}          
\end{array}
\right)
\frac{e^{ik_y y -i| q^+_x |x}}{\sqrt{A}}\  ,    
\end{equation}
 and 

\begin{equation}
\label{wave_bm_c}
\Psi^{E({\tilde Q}_-)}_{Y}( {\bf k};{\bf r}) = \  \frac{{\tilde Q}_-}{\sqrt{(1+ \text{b}^{2}_{s^{\prime}}(\pi-\theta_{q^{-}}))(1+\text{b}^{2}_{-}(\pi-\theta_{q^{-}}) )}} \, \left(
\begin{array}{c}
-\text{e}^{ i \theta_{q^{-}}} \\
\text{b}_{s^{\prime}}( \pi- \theta_{q^{-}})  \\
\text{b}_-(\pi- \theta_{q^{-}})  \\
-\text{b}_{s^{\prime}}(\pi- \theta_{q^{-}})  \text{b}_-(\pi- \theta_{q^{-}}) \text{e}^{ -i \theta_{q^-}}          
\end{array}
\right)
\frac{e^{ik_y y -i| q^-_x |x}}{\sqrt{A}}\  ,    
\end{equation}
where ${\tilde P}^{\tau}$ and ${\tilde Q}^{\tau}$ are the coefficient of forward moving and backward moving particles, $q_{\tau} = \frac{\varepsilon^{E}_Y - V_0}{s^{\prime}\hbar\upsilon_F [(1+\tau\Delta_0) + 2\tilde\zeta^2 \cos^2\theta_{q^{\tau}} (1+\tau \Delta_0^3)]}$, $q^{\tau}_{x} = q_{\tau} \cos\theta_{q^{\tau}}$, $k_{y} = k_{\tau} \sin\theta_{k^{\tau}}$ and $s^{\prime}=$sign($\varepsilon^{E}_Y - V_0$).

\medskip
\par
Using the continuity condition for the wave functions at $x=0$ and at $x=d$, we obtain  eight simultaneous  equations for eight unknowns, i.e.,  ${\tilde r}_+,  {\tilde r}_-, {\tilde t}_+, {\tilde t}_-, {\tilde P}_+, {\tilde P}_-, {\tilde Q}_+, {\tilde Q}_-$,  so that ${\tilde M} {\tilde L}={\tilde W}$ where,

\[
{\tilde M} =
\begin{pmatrix}
{\tilde M}_{11} & {\tilde M}_{12} & \cdots &{\tilde M}_{18} \\
{\tilde M}_{21} & {\tilde M}_{22} & \cdots & {\tilde M}_{28} \\
\vdots & \vdots & \ddots & \vdots \\
{\tilde M}_{81} & {\tilde M}_{82} & \cdots & {\tilde M}_{88}
\end{pmatrix}
\]

\[
M_{ij} = f(i,j,\theta, V, k)
\quad \text{for } i,j = 1,\dots,8
\]

\[
M = (M_{ij})_{8\times 8}
\]

\begin{equation}
\tilde L=  \left(
\begin{array}{cccccccc}
{\tilde r}_+&{\tilde r}_-& {\tilde P}_+& {\tilde P}_-& {\tilde Q}_+& {\tilde Q}_-&{\tilde t}_+& {\tilde t}_- 
\end{array}
\right)^T 
\end{equation} 
 and 

\begin{equation}
{\tilde W}= \frac{1}{N_i} \left(
\begin{array}{cccccccc}
-e^{-i\theta_{k^{+}}} &- \text{b}_s(\theta_{k^{+}}) & - \text{b}_+(\theta_{k^{+}})&  - \text{b}_s(\theta_{k^{+}}) \text{b}_+(\theta_{k^{+}})e^{i\theta_{k^{+}}} & 0 & 0 & 0 & 0 
\end{array}
\right)^T \   .
\end{equation}

\medskip
\par

We follow the same procedure as  we did in the absence of irradiation to solve for the unknowns ${\tilde r}_+,  {\tilde r}_-, {\tilde t}_+, {\tilde t}_-$  from the matrix equation ${\tilde M} {\tilde L}={\tilde W}$.  We have

 \begin{eqnarray}
\alpha = \frac{ (1+\text{b}^2_s(\theta_{k^+}))(1+\text{b}^2_+(\theta_{k^+}))}{(1+\text{b}^2_s(\theta_{k^-}))(1+\text{b}^2_-(\theta_{k^-}))} \left[  \frac{
\begin{array}{c}
\cos\theta_{k^{-}} \{\text{b}_s(\theta_{k^{-}})(1+\text{b}^2_-(\theta_{k^{-}}))  +\Delta_0 \text{b}_-(\theta_{k^{-}})  (1+\text{b}^2_s(\theta_{k^{-}}) \} +\\
 \tilde\zeta \{  (1-\text{b}^2_s(\theta_{k^{-}}))(1+\text{b}^2_- (\theta_{k^{-}}))+\Delta_0^2 (1+\text{b}^2_s(\theta_{k^{-}}))(1-\text{b}^2_-(\theta_{k^{-}}))  \}   
\end{array}}{
\begin{array}{c}
 \cos\theta_{k^{+}} \{\text{b}_s(\theta_{k^{+}})(1+\text{b}^2_+(\theta_{k^{+}}))  +\Delta_0 \text{b}_+(\theta_{k^{+}})  (1+\text{b}^2_s(\theta_{k^{+}}) \} +\\
 \tilde\zeta \{  (1-\text{b}^2_s(\theta_{k^{+}}))(1+\text{b}^2_+(\theta_{k^{+}}))+\Delta_0^2 (1+\text{b}^2_s(\theta_{k^{+}}))(1-\text{b}^2_+(\theta_{k^{+}}))  \}  
 \end{array}
  }\right] 
 \label{alpha}
 \end{eqnarray}

 \begin{eqnarray}
\Lambda =   \frac{(1+\text{b}^2_s(\theta_{k^+}))(1+\text{b}^2_+(\theta_{k^+}))}{(1+\text{b}^2_s(\pi-\theta_{k^+}))(1+\text{b}^2_+(\pi-\theta_{k^+}))} \left[  \frac{
\begin{array}{c}
\cos\theta_{k^{+}} \{\text{b}_s(\pi-\theta_{k^{+}})(1+\text{b}^2_+(\pi-\theta_{k^{+}}))  +\Delta_0 \text{b}_+(\pi-\theta_{k^{+}}) \times  \\
(1+\text{b}^2_s(\pi-\theta_{k^{+}}) \} -\tilde\zeta \{  (1-\text{b}^2_s(\pi-\theta_{k^{+}}))(1+\text{b}^2_+ (\pi-\theta_{k^{+}}))\\
 +\Delta_0^2 (1+\text{b}^2_s(\pi- \theta_{k^{+}}))(1-\text{b}^2_+(\pi-\theta_{k^{+}}))  \}   
\end{array}}{
\begin{array}{c}
 \cos\theta_{k^{+}} \{\text{b}_s(\theta_{k^{+}})(1+\text{b}^2_+(\theta_{k^{+}}))  +\Delta_0 \text{b}_+(\theta_{k^{+}})  (1+\text{b}^2_s(\theta_{k^{+}}) \} +\\
 \tilde\zeta \{  (1-\text{b}^2_s(\theta_{k^{+}}))(1+\text{b}^2_+(\theta_{k^{+}}))+\Delta_0^2 (1+\text{b}^2_s(\theta_{k^{+}}))(1-\text{b}^2_+(\theta_{k^{+}}))  \}  
 \end{array}
  }\right]
 \label{lambda}
 \end{eqnarray}   
  and 
 
  \begin{eqnarray}
\kappa=  \frac{(1+\text{b}^2_s(\theta_{k^+}))(1+\text{b}^2_+(\theta_{k^+}))}{(1+\text{b}^2_s(\pi-\theta_{k^-}))(1+\text{b}^2_-(\pi-\theta_{k^-}))} \left[  \frac{
\begin{array}{c}
\cos\theta_{k^{-}} \{\text{b}_s(\pi-\theta_{k^{-}})(1+\text{b}^2_-(\pi-\theta_{k^{-}}))  +\Delta_0 \text{b}_-(\pi-\theta_{k^{-}}) \times   \\
 (1+\text{b}^2_s(\pi-\theta_{k^{-}}) \}-\tilde\zeta \{  (1-\text{b}^2_s(\pi-\theta_{k^{-}}))(1+\text{b}^2_- (\pi- \theta_{k^{-}}))+\\
 \Delta_0^2 (1+\text{b}^2_s(\pi- \theta_{k^{-}}))(1-\text{b}^2_-(\pi-\theta_{k^{-}}))  \}   
\end{array}}{
\begin{array}{c}
 \cos\theta_{k^{+}} \{\text{b}_s(\theta_{k^{+}})(1+\text{b}^2_+(\theta_{k^{+}}))  +\Delta_0 \text{b}_+(\theta_{k^{+}})  (1+\text{b}^2_s(\theta_{k^{+}}) \} +\\
 \tilde\zeta \{  (1-\text{b}^2_s(\theta_{k^{+}}))(1+\text{b}^2_+(\theta_{k^{+}}))+\Delta_0^2 (1+\text{b}^2_s(\theta_{k^{+}}))(1-\text{b}^2_+(\theta_{k^{+}}))  \}  
 \end{array}
  }\right] \  . 
 \label{kappa}
 \end{eqnarray}

{51}

\end{document}